\documentclass[reprint,amsmath,amssymb,aps]{revtex4-2}
\usepackage{graphicx}
\usepackage{dcolumn}
\usepackage{bm, soul}
\usepackage[hidelinks]{hyperref}
\usepackage{float}
\usepackage{xfrac}
\usepackage{upgreek}
\usepackage{ragged2e}
\usepackage{amsmath}
\newsavebox{\tempbox}
\usepackage{qcircuit}
\usepackage{sidecap}
\usepackage{braket}
\usepackage{tikz}

\newcommand\circled[1]{\tikz[baseline=(char.base)]{
            \node[shape=circle,draw,inner sep=1.2pt] (char) {#1};}}

\begin{document}

\title{Real-time frequency estimation of a qubit without single-shot-readout}

\author{Inbar Zohar$^1$}
\email{inbar.aharon@weizmann.ac.il}
\author{Ben Haylock$^2$}
\author{Yoav Romach$^3$}
\author{Muhammad Junaid Arshad$^2$}
\author{Nir Halay$^3$}
\author{Niv Drucker$^3$}
\author{Rainer St\"ohr$^4$}
\author{Andrej Denisenko$^4$}
\author{Yonatan Cohen$^3$}
\author{Cristian Bonato$^2$}
\author{Amit Finkler$^1$}

\affiliation{$^1$Department of Chemical and Biological Physics, Weizmann Institute of Science, Rehovot 7610001, Israel}
\affiliation{$^2$Institute of Photonics and Quantum Sciences, SUPA, Heriot-Watt University, Edinburgh EH14 4AS, United Kingdom}
\affiliation{$^3$Quantum Machines, Tel-Aviv 6744332, Israel}
\affiliation{$^4$Third Institute of Physics, University of Stuttgart, Stuttgart 70569, Germany}

\date{\today}

\begin{abstract}
Quantum sensors can potentially achieve the Heisenberg limit of sensitivity over a large dynamic range using quantum algorithms. The adaptive phase estimation algorithm (PEA) is one example that was proven to achieve such high sensitivities with single-shot readout (SSR) sensors. However, using the adaptive PEA on a non-SSR sensor is not trivial due to the low contrast nature of the measurement. The standard approach to account for the averaged nature of the measurement in this PEA algorithm is to use a method based on `majority voting'. Although it is easy to implement, this method is more prone to mistakes due to noise in the measurement. To reduce these mistakes, a binomial distribution technique from a batch selection was recently shown theoretically to be superior, as all ranges of outcomes from an averaged measurement are considered. Here we apply, for the first time, real-time non-adaptive PEA on a non-SSR sensor with the binomial distribution approach. We compare the mean square error of the binomial distribution method to the majority-voting approach using the nitrogen-vacancy center in diamond at ambient conditions as a non-SSR sensor. Our results suggest that the binomial distribution approach achieves better accuracy with the same sensing times. To further shorten the sensing time, we propose an adaptive algorithm that controls the readout phase and, therefore, the measurement basis set. We show by numerical simulation that adding the adaptive protocol can further improve the accuracy in a future real-time experiment.
\end{abstract}

\maketitle

\section{Introduction}
Quantum sensing is a promising technology with many possible applications in fields such as renewable energy~\cite{Crawford2021}, condensed matter physics \cite{Sar2015, Gross2017, Dovzhenko2018, Jenkins2019}, biology \cite{Shi2015, Lovchinsky2016, Barry2016}, and chemistry \cite{SchaeferNolte2014, Finkler2021}. Different quantum systems are studied as quantum sensors \cite{Degen2017}, and depending on the systems' interactions with the environment it can be used to sense different physical quantities such as magnetic fields \cite{Jenkins2019},  electric fields \cite{Barry2016}, temperature \cite{Neumann2013}, strain \cite{Trusheim2016}, or pressure \cite{Ho2021}. One of the advantages of these sensors is the possibility of achieving high sensitivity while overcoming the standard quantum limit (SQL) and reaching the Heisenberg limit (HL) \cite{Higgins2007}.

Recent studies have pushed the sensitivity to the HL using entanglement \cite{Bollinger1996}, or quantum algorithms \cite{Vorobyov2021}. One algorithm, widely studied, is the phase estimation algorithm (PEA), suggested by Kitaev \cite{Kitaev1995}. This algorithm aims to estimate a phase $(\phi)$ that a quantum sensor is accumulating due to some interaction with frequency $f$ with an unknown parameter in the environment. The sensor accumulates the phase at $K+1$ different sensing times ($\tau$) that grow exponentially, $\tau=2^k \tau_0$, where $k$ is an index going from $0$ to $K$. The shortest sensing time $\tau_0$ limits the dynamic range (DR) of the sensor to
\begin{equation}
\mathrm{DR} = [-\frac{1}{2\tau_0},\frac{1}{2\tau_0}]
\label{eq:DR}
\end{equation}
The longest sensing time is bounded from above by the dephasing time, $T_2^*$, of the sensor $2^K \tau_0<T_2^*$ \cite{Said2011}. 

The full algorithm is based on a quantum system with multiple quantum bits that carry the process of estimating the phase simultaneously using the quantum Fourier transform \cite{Vorobyov2021}. Such multi-qubit systems are still challenging and sometimes not available for every sensing environment. Moreover, a single qubit sensor such as a spin-1/2 offers the ultimate spatial resolution, and any additional gain from entangling it with additional spins is canceled by the increase in sensor size. In these cases, therefore, a system of a single qubit that is incorporated with an adaptive PEA, and making use of quantum-classical interfaces \cite{Okamoto2012,Danilin2018}, can be of benefit. Here we experimentally demonstrate, in real-time, a non-adaptive PEA scheme in non-ideal but very realistic sensing conditions, and show numerically the advantage of moving this method to an adaptive one.

\begin{figure}[b]
\includegraphics[width=1\linewidth]{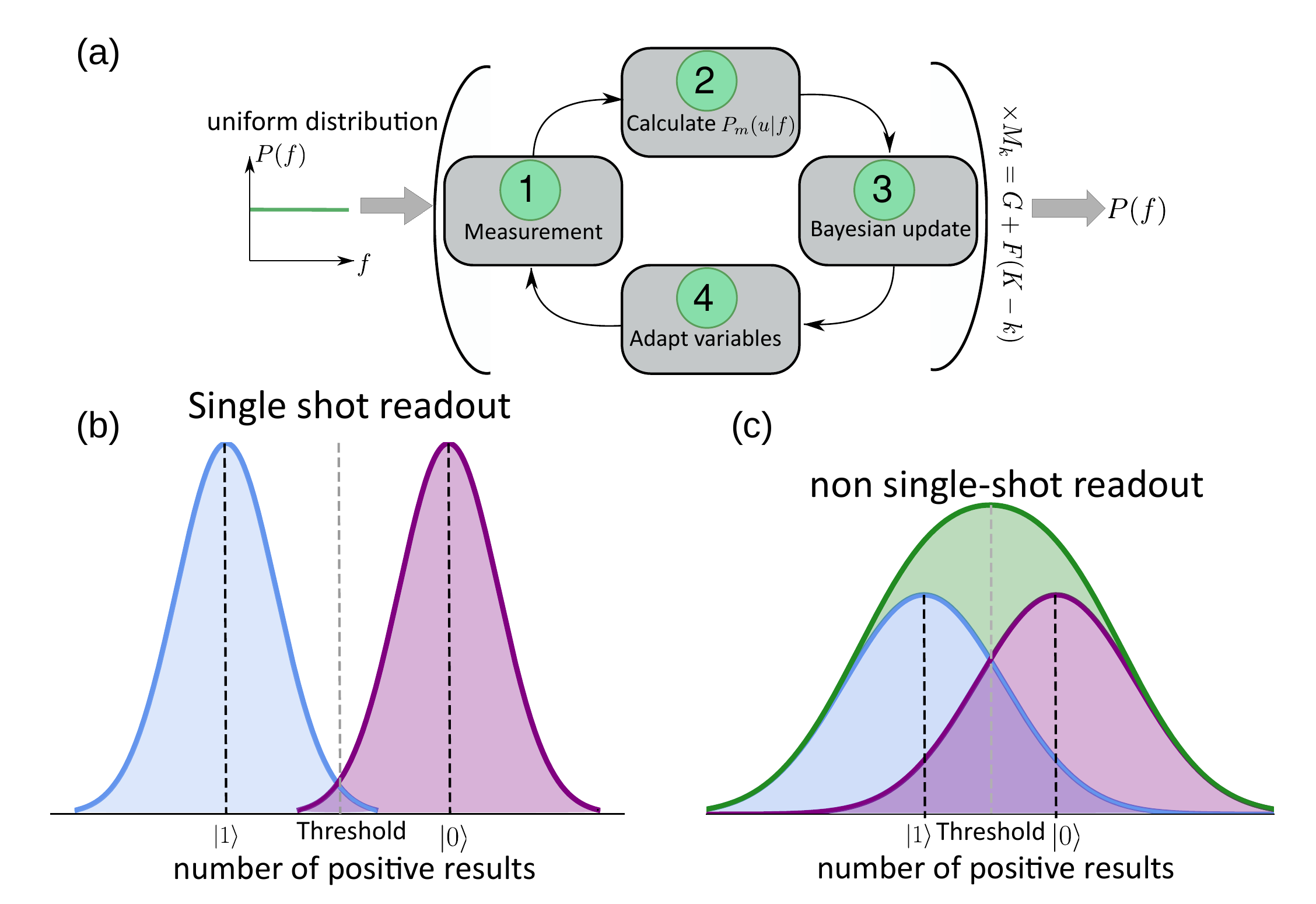}  
\caption{(a) Graphical illustration of the adaptive phase estimation algorithm comprising four steps: (1) A pulse sequence suitable for the estimation of the unknown parameter, given the nature of the interaction between the sensor and the parameter. This pulse sequence will be applied with exponentially growing sensing times. The state of the sensor is measured after every sensing time. (2) Calculating the probability function for the state of the sensor given the unknown parameter. (3) Using Bayes' Theorem to update the probability function for the parameter. (4) Calculating the optimal variables for extracting maximal information from the next iteration. After $M_k$ iterations for each sensing time, the final distribution will be the estimate of the unknown parameter. (b-c) Schematic illustration of the measurement outcome of a single-shot (b) or averaged (c) sensor.}
\label{fig1}

\end{figure}

\subsection{Adaptive phase estimation algorithm}

The general scheme for applying adaptive PEA (Fig.\,\ref{fig1}a) consists of a cyclic process of four steps. The first is a pulse sequence applied on the sensor depending on the target frequency, $f$, in question and its interaction with the sensor as expressed in the Hamiltonian, $H(f)$. This pulse sequence will use the same exponentially growing sensing time as in the quantum PEA only in sequential order, from the shortest to the longest, and not simultaneously, similar to the Kitaev's iterative PEA \cite{Kitaev2002}. After each pulse sequence with one sensing time, the sensor is measured, and the outcome ($u$) can be one of the two states of the sensor - zero or one. This outcome is used in the second step to update the probability function, $P_m(u|f)$, to measure the sensor state, $\ket{0}$ or $\ket{1}$, given that there is interaction with the unknown parameter with frequency $f$. The nature of the sensor's interaction with the target parameter in the pulse sequence is encoded in the probability function. 

The critical step of the algorithm is in step \circled{3}, where one applies a Bayesian update to estimate the unknown parameter \cite{Bonato2015, Valeri2020, Gebhart2022}
\begin{equation}
P_\mathrm{posterior} (f|u)\propto P_m (u|f) P_\mathrm{prior} (f|u)
\label{eq:Bayesian}
\end{equation}
where $P(f|u)$ is the probability function of the measurement outcome given the target parameter, subscript \texttt{posterior} is the new probability after each Bayesian update and \texttt{prior} is the old one from the last update. $P(u|f)$ is the probability function of the target parameter given the outcome of the measurement is $u$, the subscript $m$ denotes a single outcome. Since the adaptive PEA applies the sensing scheme with different sensing times sequentially, each measurement holds less information about the phase than the quantum PEA. The penalty in the full scheme is that each sensing time is measured multiple times by changing one of the sensing variables, as is illustrated in step \circled{4}. The number of iterations $M_k=G+(K-k)F$ for each sensing time grows as the sensing time gets shorter, where $G$ and $F$ are optimized parameters, and $k$ is the index of the sensing time \cite{Higgins2009}. The adaptive character of the scheme is established in step \circled{4}. In this step, the optimal variables for gaining maximal information are calculated based on the last probability function and then transferred to the pulse sequence of the next iteration. 

Adaptive PEA has been studied extensively. Theoretical works suggested controlling the sensing phase or the sensing time \cite{Cappellaro2012} to enhance sensitivity. Others used numerical simulations \cite{Wiebe2016, Scerri2020}, and several did experimental studies with different sensors \cite{Santagati2019, Bonato2015} to prove the feasibility and benefits of this protocol. All of these studies were performed with a single-shot readout (SSR) sensor, where the state of the sensor can be measured after one measurement with high fidelity (Fig.\,\ref{fig1}b). Nevertheless, in some cases, non-SSR sensors are the only possible sensing approach, for instance, for imaging nanoscale biological samples  with high special resolution and in ambient conditions. These sensors are characterized by the high ratio of classical noise added in the measurement, for example, low photon collection efficiency in optically read-out systems, compared to the quantum projection noise of the system  \cite{Degen2017}. This causes the histogram of the measurement outcomes to mix `0' and `1'. Therefore, assigning the measurement outcome to one state of the sensor with high fidelity, i.e., in one shot, is impossible (Fig.\,\ref{fig1}c).

For a non-SSR sensor, the pulse-sequence and the measurement should be applied for many repetitions to assess the sensor state, still with a non-negligible error. This situation requires adjusting the probability function $P(u|f)$ used in the Bayesian update to the averaged measurement result. The most common and simple solution is to use a threshold that is calculated based on the probability to measure a positive outcome from the sensor at each of the states, which can be a collection of photons for an optically measurable sensor (See Appendix \ref{apx:visibility}). In this method the measurement is repeated for $R$ times and the number of positive outcomes $r$ is assigned to a state of the sensor, $u$, based on the calculated threshold; we call this method ``majority voting''. This approach results in a binary outcome from a large batch of size $R$ repetitions of the measurement. This method's benefit is the possibility of using the probability function and Bayesian update as in the SSR sensor scheme \cite{Joas2021}. However, it disregards most of the possible outcomes from the $R$ repetitions by using only a binary span of results. Therefore, it is also more prone to noise, where a noisy measurement can be assigned to the wrong binary option \cite{Dinani2019}.

Since a non-SSR measurement entails repeating the measurement $R$ times to improve the readout fidelity, we consider the number of positive outcomes, $r$, out of the full $R$ batch. This probability distribution, then, is binomial,
\begin{equation}
P(f|r)=\left(
\begin{array}{c}
{R}\\{r}
\end{array}\right)
P_d(1|f)^r[1-P_d(1|f)]^{R-r}
\label{eq:binomial}
\end{equation}
where $P_d(1|f)$ is the probability of detecting a positive outcome given the sensor state was '1' calculated for the full range of the unknown parameter $f$. The subscript $d$ denotes the detection of the positive outcome, and $r$ is the number of positive outcomes of the measurement \cite{Dinani2019}. In this case, the information about the phase accumulated due to the external target parameter is encoded in $P_d$. This method considers the full range of possible outcomes for the averaged measurement. Therefore, a noisy measurement will not result in a mistake but with an error within the range of the noise of the measurement - and therefore leading to a more sensitive estimation \cite{Dinani2019}. So far, experimental demonstration of the binomial distribution approach and the enhancement in accuracy it offers has not been demonstrated.

\subsection{DC magnetometry with non-SSR sensor}

\begin{figure*}[t]
  \includegraphics[width=1\linewidth]{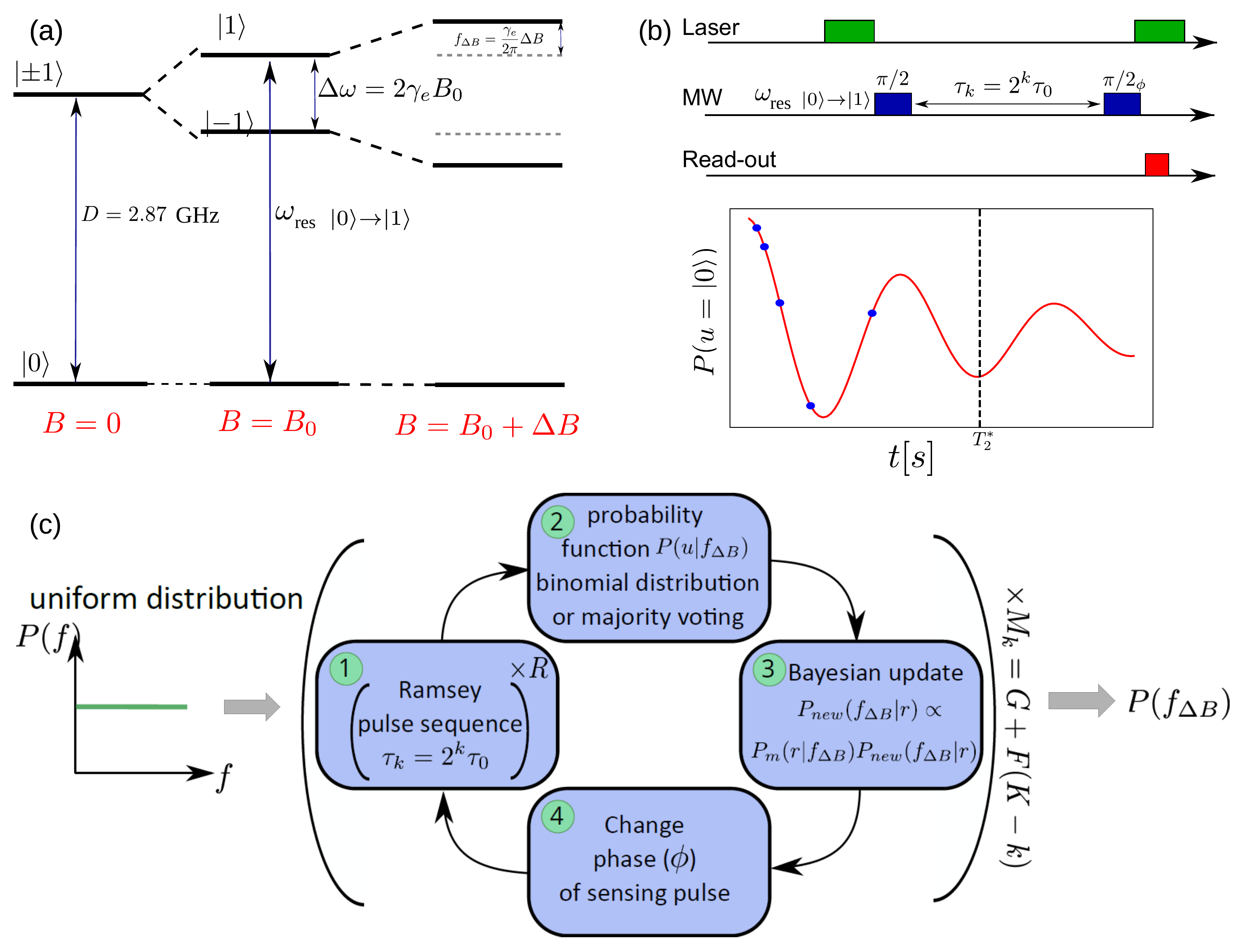}
\caption{(a) Illustration of the energy levels of the NV center under a static magnetic field with a small detuning. The errors indicate the microwave field resonant with the $\ket{0} \rightarrow \ket{1}$ transition. (b) The top panel is a Ramsey pulse sequence applied on an NV center for sensing a DC magnetic field. The red curve is a simulated result for a Ramsey sequence with a logarithmic array of sensing times between zero and $T_2^*$. The blue points indicate the $K$ sensing times used in the PEA scheme. (c) Graphical illustration of adaptive phase estimation algorithm for DC magnetometry with a non-SSR sensor.}
\label{fig2}
\end{figure*}

In this work, we use the nitrogen-vacancy (NV) center in diamond, a widely used non-SSR sensor at ambient conditions \cite{Maze2008, Balasubramanian2008}. The NV center is a spin-1 system with a zero field energy splitting between $m_s=0$ and $m_s=\pm 1$ spin states of $2.87$\,GHz (implicitly $\hbar$ is taken to be equal to 1). It is sensitive to DC magnetic fields due to the Zeeman effect $H(B)=\gamma_e \textbf{B}\cdot \textbf{S}$, where $\gamma_e$ is the electron gyromagnetic ratio and $\textbf{S}$ is the spin operator. When the magnetic field is aligned with the $z$ axis of the spin, the Hamiltonian of the system is simplified to  
\begin{equation}
H_\mathrm{NV}(B)=DS_z^2+\gamma _eB_0S_z.
\label{eq:Hamiltonian}
\end{equation}
This interaction results in an energy splitting between the two ($m_s=\pm 1$) degenerate spin levels of $\Delta \omega = 2\gamma_e B$. Under these conditions and in most instances, each of the two single quantum transitions of the NV center can be practically considered as a two-level spin system (Fig.\,\ref{fig2}a).

The first step in the PEA scheme is applying the pulse-sequence sensitive to the target field (Fig.\,\ref{fig2}c). For a DC magnetic field, we use Ramsey interferometry (Ref.\,\onlinecite{Childress2006}, Fig.\,\ref{fig2}b). The evolution of the spin in such a pulse-sequence can be simplified when considered in the rotating frame. After initialization to $\ket{0}$, to prepare the sensor, a $\frac{\pi}{2}$ microwave pulse resonant with the eigenenergy of the sensor $\Delta \omega$ is applied, placing the sensor in a superposition state of the two eigenstates $\ket{\psi(t=0) }= 2^{-1/2}\left(\ket{0} + \ket{1}\right)$. When the sensor interacts with a small external magnetic field $\Delta B$, it will accumulate a relative phase between the two eigenstates in the rotating frame, which is proportional to the external magnetic field $\ket{\psi (t)}=2^{-1/2}\left(\ket{0} +e^{-i\gamma_e\Delta Bt}\ket{1}\right)=2^{-1/2}\left(\ket{0} +e^{-i2\pi f_{\Delta B}t}\ket{1}\right)$, where $f_{\Delta B}$ can also be considered as a small frequency detuning from the resonance frequency of the sensor (as illustrated in Fig.\,\ref{fig2}a), and therefore, throughout in the manuscript we use the two terms interchangeably. When applying another $\frac{\pi}{2}$ microwave pulse, we project the spin to the eigenstates of $\sigma_z$. If this pulse is rotated by an angle $\phi$ from the preparation pulse, we will project the sensor to a rotated spin basis $\sigma_xe^{-i\phi}$, where $\phi$ is the phase we change in the fourth step of the scheme in figure \ref{fig2}c. This pulse sequence can estimate external magnetic fields that are within the dynamic range of the measurement (Eq.\,\ref{eq:DR}).

The second step of the PEA (Fig.\,\ref{fig2}c) is to calculate the probability function according to the prior-measured state of the sensor $u=\ket{0} / \ket{1}$. This probability function is based on the Ramsey interferometry model for sensing a small external magnetic field $\Delta B$ (Fig.\,\ref{fig2}b),
\begin{equation}
P_m(u|\Delta B)=\frac{1}{2}\left[1+(-1)^u e^{-(\sfrac{t}{T_2^*})^2}\cos(2\pi f_{\Delta B}t-\phi)\right],
\label{eq:majority_P}
\end{equation}
where $T_2^*$ is the dephasing time of the sensor. This probability function can be used for an SSR sensor, like the NV center at cryogenic conditions \cite{Robledo2011}, or with the majority voting approach for the non-SSR sensor, like the NV center at ambient conditions \cite{Joas2021}. 

However, for the binomial approach we want to use Eq.\,\ref{eq:binomial}, which accounts for all $r$ possible outcome of the repeated measurement. This probability function depends on the probability of detecting a positive outcome given the external target parameter $P_d(1|f)$. As shown in the theoretical derivation from Ref.\,\onlinecite{Dinani2019}, this probability for sensing an external DC magnetic field is
\begin{equation}
P_d(1|f)=\alpha \left[
1-Ve^{-(\sfrac{t}{T_2^*})^2}\cos(2\pi f_{\Delta B}t -\phi)
\right]
\label{eq:binomial_gauss}
\end{equation}
where $\alpha$ is the sensor's threshold, and $V$ is the visibility of the sensor (See Appendix \ref{apx:visibility}).

\section{Real-time Bayesian update comparison}

We report for the first time on the advantage of the binomial approach over the majority voting in a real-time experiment. The experiment was done  at ambient conditions setup using a QM OPX to conduct real-time calculations (See Appendix \ref{apx:Set-up}). We collected data from a single NV center with a dephasing time of $T_2^*=3.5$\,$\upmu$s (See Appendix \ref{apx:Sample}). We used five sensing times ($K=4$) with $\tau_0=100$\,ns in the Ramsey measurement pulse sequence (First step in Fig.\,\ref{fig2}c). The probability function estimating the external magnetic field was constructed with a resolution (binning) of $25$\,kHz. For each external magnetic field, we applied the scheme twice, once with the majority voting probability function (Eq.\,\ref{eq:majority_P}) in the second step and once with the binomial distribution probability function (Eq.\,\ref{eq:binomial}). After the Bayesian update (third step in Fig.\,\ref{fig2}c), we change the phase of the second $\frac{\pi}{2}$ pulse linearly between zero and $\pi$ in a non-adaptive manner following a predetermined measurement sequence \cite{Higgins2009} (fourth step in Fig.\,\ref{fig2}c).

\begin{figure}[t]
\includegraphics[width=.975\linewidth]{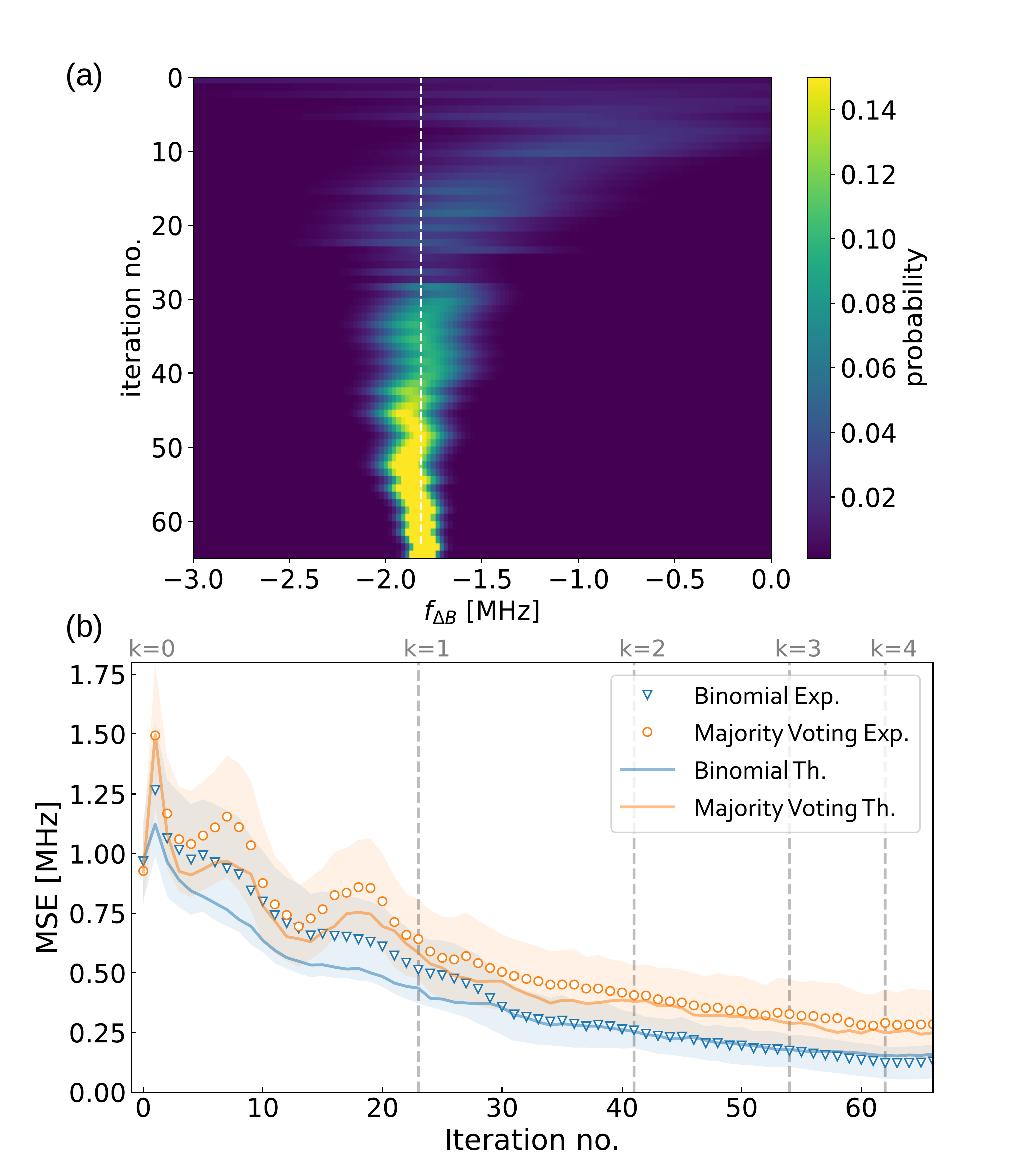}
\caption{(a) The probability function of a single magnetic field detuning as it is updated with the iterations of the scheme. The detuning here was 1.895\,MHz. (b) Error (square root of variance) as a function of total measurement time (number of iterations) with binomial distribution approach (blue) and majority voting approach (orange) for a repetition number $R=2500$. The data presented here is for 500 random chosen detunings. The vertical lines (different values of $k$) represent the move to the next $\tau$ in the algorithm.}
\label{fig3}
\end{figure}

Figure\,\ref{fig3}a presents the iteration number of a measurement of a random magnetic field using the approach described above (Bayesian, non-adaptive). The probability function starts as a uniform distribution. The first iterations apply the shortest sensing time, which guides the probability function to a rough estimation of the frequency. As the iterations advance, the sensing time gets longer, and the estimated frequency gets focused and narrower to a more precise estimate. The frequency at the peak of the probability function in the last iteration is the final estimation for this measurement. 

We applied the two approaches for a non-SSR sensor in a non-adaptive scheme on 500 randomly chosen magnetic fields $f_{\Delta B}$ in the range [-2, 2]\,MHz. For more information about the choice of the range, see Appendix \ref{apx:detunings}. We applied the external magnetic field as an off-resonance microwave tone relative to the $\ket{0}\rightarrow\ket{-1}$ transition of the NV, $\omega_{-1}$, at an applied magnetic field of 551\,Gauss (close to the NV's excited-state level anti-crossing), corresponding to $\omega_{-1} = 2\pi\times$1.3322 GHz. All 500 magnetic fields were measured with seven different repetition numbers $R=(100, 250, 500, 750, 1000, 2500, 5000)$. For each detuning we measured the two approaches in a random order and with all repetition numbers also randomized in the order. After each detuning we refocused the frequency and position of the confocal setup. 

To compare the two sensing methods, we calculated the mean square error (MSE):
$$\mathrm{MSE} = \sqrt{V_B}=\sqrt{\left< \left(\tilde{f}_B-f_B\right)^2\right>}.$$ based on the estimated frequency  $\tilde{f}_B$ calculated from the $P(f|r)$ after every iteration. Our results show a reduction of the MSE with the same measurement time when using the binomial distribution approach. The best MSE achieved was $\approx 0.12\,\mathrm{MHz}$
for $R=2500$ with a total sensing time of $T=1.07$\,s when using the binomial distribution method. The majority voting method reached a MSE of $\approx 0.28\,\mathrm{MHz}$ within the same time, more than twice as high. The lowest possible MSE is limited by the decoherence time of the sensor where $\mathrm{MSE}\geq \sqrt{\frac{1}{T_2^*}}$

We note that the MSE for larger $R$ does not improve by much, and we attribute this to the slight improvement of the contrast (See Appendix \ref{apx:contrast}). The superiority of the binomial distribution approach is evident also for shorter sensing times, starting from $R=250$ with $T=0.34$ s (see Fig.\,\ref{fig:contrast} in Appendix \ref{apx:contrast}). To see the improvement in MSE, we plot it as a function of the iteration number for the $R = 2500$ case in Fig.\,\ref{fig3}b. It improves as the iterations progress due to significant improvement in the estimation precision, smaller MSE, compared to the addition of the total sensing time needed for this improvement. We see a good agreement between the experiment and a simulation based on the experimental parameters used in the experiment. The small discrepancy between experiment and simulation can be explained by a little uncertainty in the detection probability for the 0 and 1 state, i.e., $P_d(1|m_0)$ and $P_d(1|m_1)$. The MSEs calculated for each number of repetitions, $R$, are plotted as a function of the contrast, averaged over all 500 frequencies with the same number of repetitions in Fig.\,\ref{fig:contrast} in Appendix \ref{apx:contrast}.

\section{Adaptive Bayesian update with binomial distribution}
\label{adaptive}
\begin{figure}[b]
\includegraphics[width=0.49\textwidth]{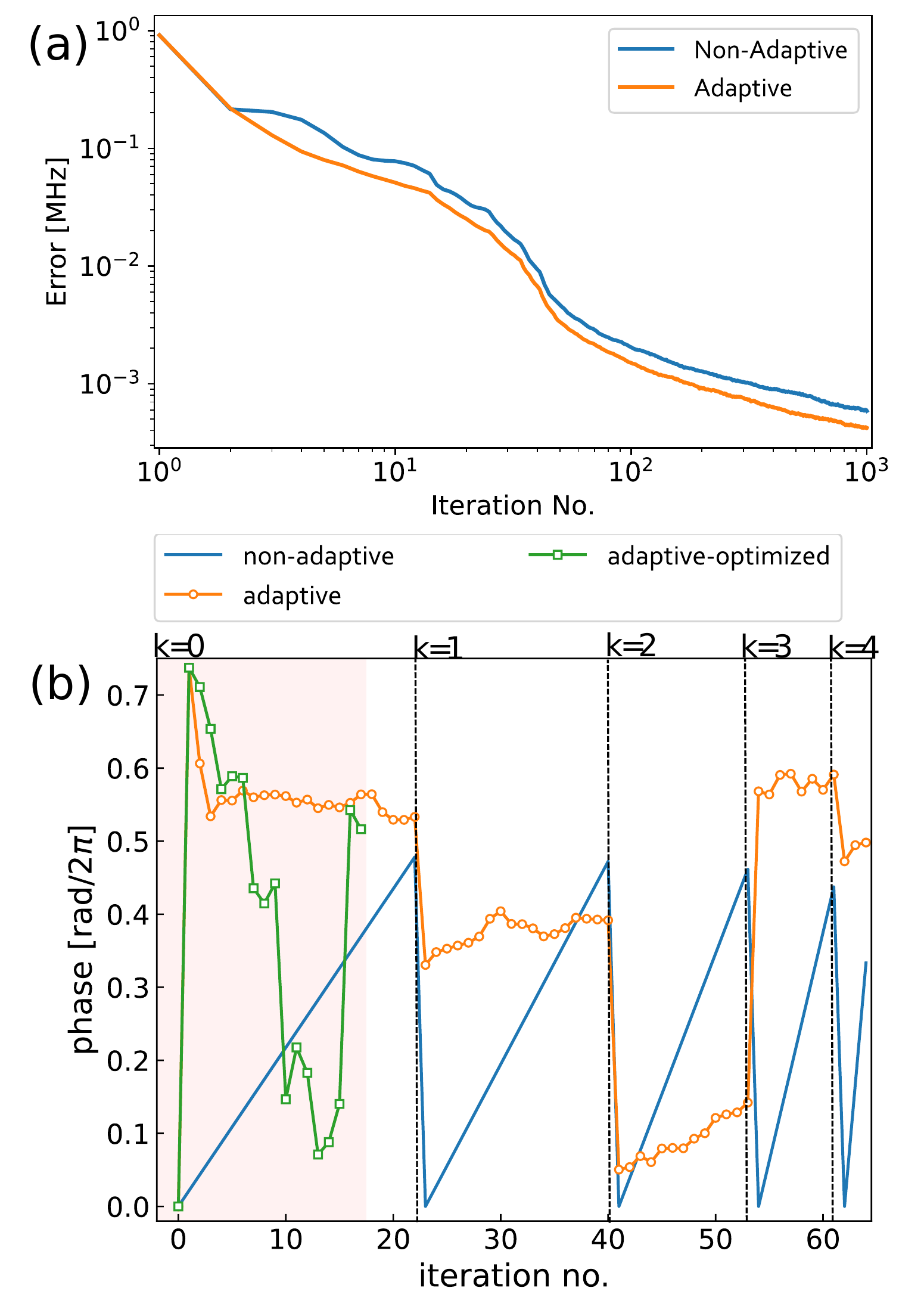}
\caption{(a) Simulated experiment with the calculated phases for the different methods: non-adaptive (blue) and our adaptive protocol (orange). After reaching the end of the settings determined by the PEA scheme, measurement times $\tau = T_2^*$, and phases are chosen at random (non-adaptive) or via adaptive algorithm. (b) Comparison between three different phase calculations for a single experiment with detuning of $f_{\Delta B}=-1.3\,\mathrm{MHz}$: non-adaptive (blue), our adaptive (orange) and the adaptive-optimized (green), the latter showing an improvement in the number of iterations (=time) needed to attain the correct phase (see main text). The vertical lines (different values of $k$) represent the move to the next $\tau$ in the algorithm.}
\label{fig4}
\end{figure}

As shown with an SSR sensor, using an adaptive scheme where the measurement variables are optimized based on the updated probability function further improves the sensitivity of the method \cite{Bonato2015}. In DC magnetometry we look for the optimal readout phase $\phi$. While one can optimize quantities such as the information gain, this is typically quite complex and adds significant computational overhead. Simpler adaptive rules can be obtained through the Cram\'er-Rao lower bound (CRLB), which represents the minimum reachable variance for any (unbiased)
estimator of $\phi$. As the CRLB of $\phi$ is inversely proportional to the Fisher information $\mathcal{I}$, one can target the maximization of $\mathcal{I}$ to improve the estimate of $\phi$. To find this phase we calculate the Fisher information of the probability function as it is written in Eq.\,\ref{eq:Information}a and maximize it with respect to the phase $\phi$,
\begin{subequations}
\begin{equation}
{\cal I}(f_{\Delta B})=E\left[\left(\frac{\partial}{\partial f_{\Delta B}}\log\left(P(r|f_{\Delta B})\right)\right)^2\right],
\end{equation}
\begin{equation}
\frac{\partial}{\partial\phi}{\cal I}(f_{\Delta B})=0.
\end{equation}
\label{eq:Information}
\end{subequations}
where $E$ is the expectation value. 

By solving the optimum problem for the phase and taking the solution that results with the maximum, we find the optimal phase,
\begin{equation}
\phi_\mathrm{opt}=2\pi E\left[f_{\Delta B}\right]t-\cos^{-1}\left(\frac{-B}{A}\right),
\label{eq:phi_opt}
\end{equation}

where $A=\frac{r^2}{R^2}+\left(1-2\frac{r}{R}\right)\alpha$ and $B=\left(1-2\frac{r}{R}\right)\alpha V$ (See Appendix \ref{apx:Adaptive}). Using the number of positive results and the expectation value at each iteration with this optimal phase calculation will result in the next readout phase. 

To evaluate the benefit of the adaptive scheme compared to the non-adaptive one, we simulated the experiments based on the dephasing time ($T_2^*$) and threshold ($\alpha$) (see Appendix \ref{apx:Sample}) of the NV used in the real-time experiment. Simulations were performed by numerically reproducing the experiment, randomly generating a simulated photon number $r$ from a binomial distribution as in Eq.\,\ref{eq:binomial_gauss}, using experimental parameter ($R=10^5$, $G=3$, $F=2$), and are presented in Fig.\,\ref{fig4}a. We observe two different regimes. The first one where we increase the $\tau$ exponentially, which is the `high dynamic range' regime. Once we reach $T_2^*$ we measure from that iteration number at $\tau = T_2^*$, at the SQL, hence the clear change in slope at iteration number (approximately) 30.  In both cases (non-adaptive and adaptive), the probability function was calculated based on the binomial distribution approach. In the simulation of the adaptive scheme, the phase of the Ramsey readout pulse in the next iteration was calculated based on the probability function (step 4 in Fig.\,\ref{fig2}c) and Eq.\,\ref{eq:phi_opt}, whereas in the non-adaptive scheme the phase was linearly ramped between zero and $\pi$ (see Fig.\,\ref{fig4}b). 

The phases calculated for the adaptive simulated experiment show convergence of the phase in a small number of iterations, smaller than $M_k$ iterations determined theoretically for each sensing time (Fig.\,\ref{fig4}b, orange circles). This convergence raises the possibility of reducing the number of iterations for each sensing time by moving to the next sensing phase once the phase remains steady for three iterations with an error of $\frac{0.1}{\pi}$ (Fig.\,\ref{fig4}b, green squares). We denote this method as adaptive-optimized. It has the potential to reduce the total measurement time significantly, which will also improve the sensitivity.

The MSE calculated from simulated results of the two methods, non-adaptive and adaptive, is plotted in Fig.\,\ref{fig4}a as a function of increasing iterations, where each iteration consists of a new value for the phase. Both methods in the simulation used the binomial distribution approach for the probability function to calculate the final estimation, as this approach proved to be more sensitive also in the real-time experiment.

\section{Conclusions}

We performed a real-time Bayesian update with an NV center, a non-SSR sensor at ambient conditions. We compared the MSE of the sensor between two calculation methods - majority voting and binomial distribution, and showed that the latter approach has better sensitivity than the former.

We showed by simulation that an adaptive scheme can further improve the MSE, and suggested using it to also reduce the total number of iterations and therefore the total sensing time, and offer extra improvement on the sensitivity. Our simulations suggest that these schemes can achieve a sensitivity four times better than the non-adaptive approach.

This work demonstrates how one can use non-SSR sensors as practical tools in adaptive PEA, and serves as a proof of concept for a specific non-SSR sensor, the NV center in diamond. Nevertheless, it could also be implemented in other sensing systems, as the approach is general. This method can also be used in other sensing schemes, such as ac magnetometry using dynamical decoupling \cite{Staudacher2013} for solid state spin sensors. 

\paragraph*{Acknowledgements.} We thank B.\,Nadler for invigorating discussions. C.\,B.\, and A.\,F.\, are jointly supported by the ``Making Connections'' Weizmann-UK program. C.\,B.\, is supported by the Engineering and Physical Sciences Research Council (EP/S000550/1 and EP/V053779/1). A.\,F.\, acknowledges financial support from the Israel Science Foundation (ISF Grant No.\,963/19). A.\,F.\, is the incumbent of the Elaine Blond Career Development Chair in Perpetuity and acknowledges the historic generosity of the Harold Perlman Family, research grants from the Abramson Family Center for Young Scientists and the Willner Family Leadership Institute for the Weizmann Institute of Science.

\section{Appendices}
\renewcommand\thefigure{A\arabic{figure}}    
\subsection{Sample}
The NV layer was created by a 10\,keV nitrogen ion ($^{15}\mathrm{N}^+$) implantation with a flux of 80 ions per $\upmu\mathrm{m}^2$ in an electronic grade (e6) CVD diamond, subsequently annealed in vacuum at a temperature of 950\,$^\circ\mathrm{C}$ for two hours. A nanopillar structure was then etched in the diamond for enhanced photon collection efficiency \cite{Momenzadeh2015}. All measurements were performed on a single NV center. The dephasing time of the NV center is $T_2^* = 3.5\,\upmu$s, measured with a standard Ramsey (FID) sequence on resonance. The Rabi contrast of the NV center was about ~30\%\ with count rate of 80 kcounts per second. 
\label{apx:Sample}

\subsection{Experimental setup}
The NV center was measured on a custom-built (confocal microscope with a 520 nm laser diode for excitation, dichroic mirror for separating excitation and fluorescence, a band-pass filter for fluorescence counting and two avalanche photodiodes in a Hanbury-Brown and Twiss configuration). We used a QM OPX to orchestrate all pulse sequence generation, photon readout, real-time Bayesian estimation and adaptive phase calculation. A local oscillator from a Windfreak SynthNV-Pro was mixed (Marki MLIQ-0218L) with two low-frequency (150\,MHz) 90$^\circ$ phase-shifted sine signals from the OPX to produce a single-sideband modulated RF, amplified by an EliteRF (M.02006G424550) broadband amplifier.

As opposed to prior works with NVs \cite{Santagati2019, Joas2021, McMichael2021}, here the measurements and Bayesian update are performed by an FPGA-based computer in real-time (QM OPX). Together with on-the-fly pulse sequence generation, each Bayesian Update (in the non-adaptive case) takes only 0.4\,ms to complete (for a probability distribution function of length 400, or 1\,$\upmu\mathrm{s}$ per frequency bin), with a small overhead of $<1\,\upmu\mathrm{s}$ for the optimal phase calculation (in the adaptive case, Eq.\,\ref{eq:phi_opt}). 
The OPX QUA code is available on \href{https://github.com/hellbourne/be-paper}{github}.

Measurements for Fig.\,\ref{fig3}b were taken on a similar setup, previously described by Arshad et al. \cite{Arshad2022}, using a laser-written NV center with $T_2^* = 5.5\,\upmu$s. A single photon count rate of $\approx 50$\,kcps was equivalent to $P_d(1|m_1) \approx 0.011$ and $P_d(1|m_0) \approx 0.016$.
\label{apx:Set-up}
\newpage
\onecolumngrid
\subsection{Adaptive phase calculation}
Taking Eq.\,\ref{eq:binomial} and the Ramsey model (Eq.\,\ref{eq:binomial_gauss}), we derive in this appendix the optimal phase in the adaptive case. First, define the model as $L(f_B, \theta) = \alpha\left[1+V\cos\left(2\pi f_B\tau - \theta\right)\right]$ and calculate its derivative
$$
L'(f_B, \theta) = \frac{\partial}{\partial f_B} L(f_B, \theta) = -2\alpha V \pi \tau \sin(2\pi f_B\tau-\theta).
$$
Next, we use the binomial probability distribution to write down the mean and variance,
\begin{align*}
    \mu_r & = E\left[r | f_B\right] = R\cdot L(f_B, \theta)\\
    \sigma_r^2 & = E\left[\left(r-\mu_r\right)^2 | f_B\right] = R\cdot L(f_B, \theta)\left[1-L(f_B,\theta)\right].
\end{align*}
We can approximate $L(f_B,\theta) \equiv L \simeq \frac{r}{R} + \Delta$ if $\Delta \ll 1$, and then get an expression for the variance in leading orders of $\frac{r}{R}$:
\begin{align*}
    \sigma^2_r & = R\cdot L(f_B, \theta)\left[1-L(f_B,\theta)\right] = RL-RL^2 = R\left[\left(\frac{r}{R} + \Delta\right) - \left(\frac{r}{R} + \Delta\right)^2\right] =  \\
    & = R\left[ \left(\frac{r}{R} + \Delta\right) - \left(\frac{r^2}{R^2} + 2\Delta\frac{r}{R} + \Delta^2\right)  \right] \approx R\left[\left(\frac{r}{R} + \Delta\right) - \frac{r^2}{R^2} - 2\Delta\frac{r}{R}\right] = \\
    & = R\left[ L-\frac{r^2}{R^2} - 2\left(L-\frac{r}{R}\right)\frac{r}{R}  \right] = R\left[ L + \frac{r^2}{R^2} - 2\frac{r}{R}L \right] = R\left[L\left(1-2\frac{r}{R}\right) + \frac{r^2}{R^2}\right]
\end{align*}
Now we define the logarithm of the model (likelihood) function:
\begin{align*}
    K(r, f_B) & \equiv \log P(r | f_B) = \log \left(
\begin{array}{c}
{R}\\{r}
\end{array}\right) + r\log L(f_B,\theta) + (R-r)\log\left(1-L(f_B,\theta)\right)
\end{align*}
and so
\begin{align*}
        \frac{\partial}{\partial f_B}K(r, f_B) & = \frac{rL'}{L} - \frac{(R-r)L'}{1-L} = \left[ \frac{r}{L} - \frac{R-r}{1-L} \right]L' = \left[ \frac{r-RL}{L(1-L)}L'   \right] = \frac{R}{\sigma_r^2}(r-\mu_r)L'(f_B, \theta).
\end{align*}
As we wrote in Sec.\,\ref{adaptive}, the Fisher information can now be explicitly calculated,
\begin{align*}
    I(f_B) & = E\left[  \left(\frac{\partial}{\partial f_B}K(r,f_B)\right)^2\right] = E\left[(r-\mu_r)^2\right]\frac{R^2}{\sigma_r^4}\left(L'(f_B,\theta)\right)^2 = \frac{R^2}{\sigma_r^2}\left(L'(f_B,\theta)\right)^2 \\ 
    & \approx \frac{R\left(L'(f_B,\theta)\right)^2}{\frac{r^2}{R^2} + \left(1-2\frac{r}{R}\right)L(f_B,\theta)} = 4R\alpha^2 V^2\pi^2\tau^2\frac{\sin^2\left(2\pi f_B\tau - \theta\right)}{\frac{r^2}{R^2} + \left(1-2\frac{r}{R}\right)\alpha\left[1+V\cos\left(2\pi f_B\tau - \theta\right)\right]}
\end{align*}
The last term can be written in a more compact form by denoting
\begin{align*}
A & = \frac{r^2}{R^2} + \left(1-2\frac{r}{R}\right)\alpha \\
B & = \left(1-2\frac{r}{R}\right)\alpha V \\
C & = 4R\alpha^2V^2\pi^2\tau^2,
\end{align*}
such that,
$$
I(f_B) = C\frac{\sin^2(2\pi f_B\tau-\theta)}{A+B\cos(2\pi f_B\tau-\theta)}
$$
We maximize the Fisher information (or minimize the Cram\'er-Rao bound):
$$
\frac{\partial}{\partial \theta} I(f_B) = 0
$$
with two solutions. The first one is a minimum with $\theta = 2\pi f_B \tau$ and the second solution is $ A\cos(2\pi f_B\tau - \theta) + B = 0$. Using $\hat{f}_B = E[f_B]$, gives:
$$
\theta_\mathrm{opt} = 2\pi \hat{f}_B \tau - \cos^{-1}\left( \frac{-B}{A} \right),
$$
\label{apx:Adaptive}
\twocolumngrid

\subsection{Contrast}
As defined previously \cite{Bonato2015}, the contrast, $C$, for $R$ repetitions scales as 
$$
    C = \left[ 1+ \frac{2(\alpha_0 + \alpha_1)}{(\alpha_0 - \alpha_1)^2R}\right]^{-1/2}
$$
where $\alpha_0$ is the number of photons per shot when the NV is in the $m_s=0$ state and $\alpha_1$ is the number of photons per shot when the NV is in the $m_s = 1$ state. In Fig.\,\ref{fig:contrast} we plot the MSE as a function of the contrast, $C$:
\begin{figure}[ht]
    \centering
    \includegraphics[width = 0.5\textwidth]{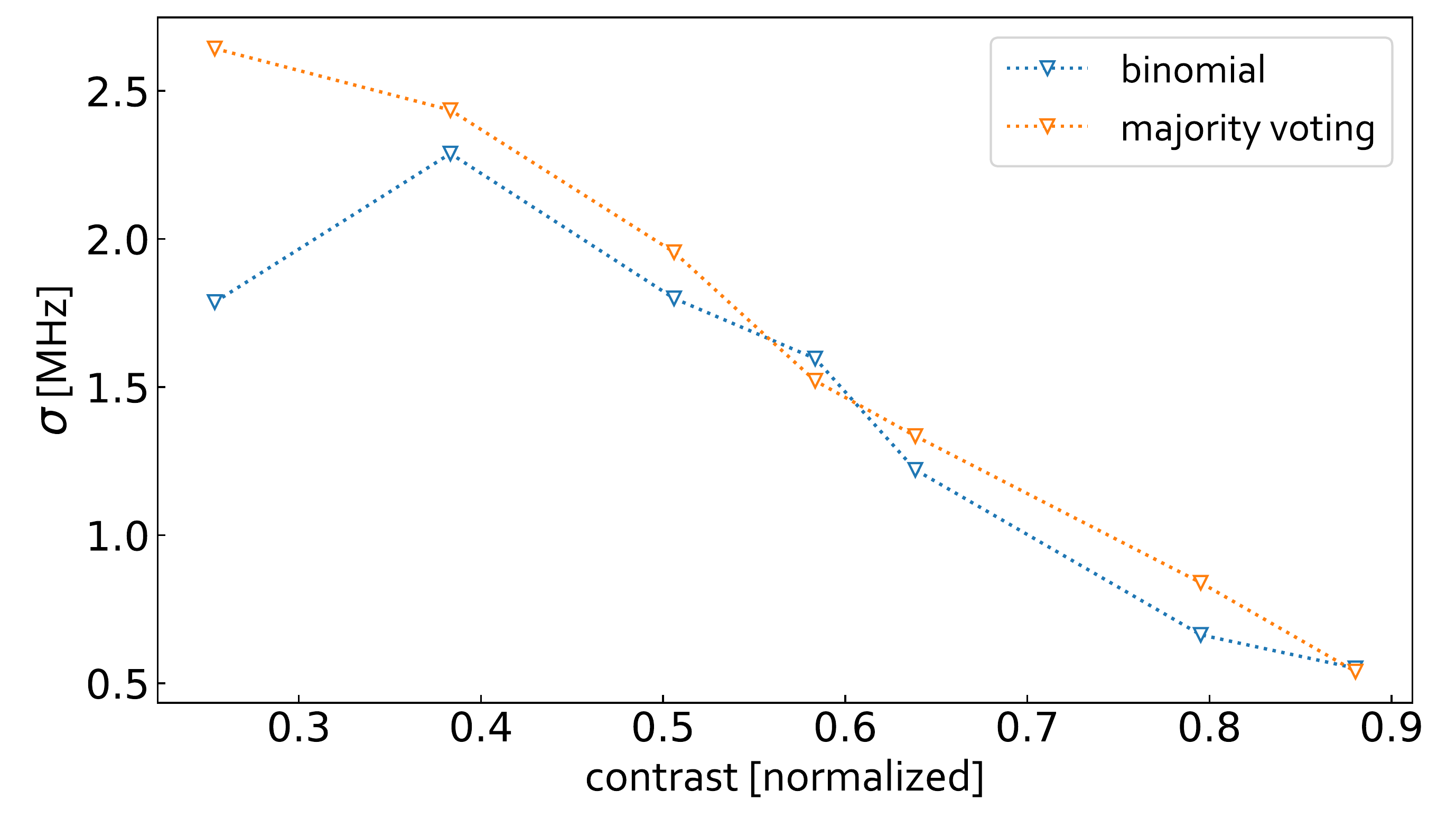}
    \caption{MSE as function of contrast. For typical values of $\alpha_0$ and $\alpha_1$ from our experiment and our choice of different $R$ in the range of $[100,5000]$, this translates to a contrast range from nearly zero to 0.9.}
    \label{fig:contrast}
\end{figure}
The data presented in this figure is from a different dataset than that plotted in Fig.\,\ref{fig3}b, hence the difference in the MSE. Nevertheless, the point regarding the contrast is still valid.
\label{apx:contrast}

\subsection{Visibility}
In Eq.\,\ref{eq:binomial_gauss} we introduced two parameters: the threshold, $\alpha$ and the visibility, $V$. Following Ref.\,\onlinecite{Dinani2019} and for self-consistency, we define them here as
\begin{align*}
    \alpha & = \frac{1}{2}\left[ P_d(1|m_0) + P_d(1|m_1)\right] \\
    V      & = \frac{P_d(1|m_0) - P_d(1|m_1)}{P_d(1|m_0) + P_d(1|m_1)}e^{(\tau/T_2^*)^2},
\end{align*}
where $P_d(1|m_i)$, for $i=0,1$, is the probability of a detector click for the spin in the state $\ket{i}$. Typical values for our setup were $P_d(1|m_1) = 0.0251$ and $P_d(1|m_0) = 0.03419$.
\label{apx:visibility}

\subsection{Choice of range for random detunings}
As we explained in the main text, the dynamic range of the sensor is bounded in the range $\left[ -\tfrac{1}{2\tau_0}, \tfrac{1}{2\tau_0}\right]$ which is $[-5, 5]$\,MHz in our case. While we saw no discernible change in the conventional Ramsey curve for detunings larger than 2\,MHz, the level of noise increased dramatically for the Bayesian update. We therefore limited the range to $[-2,2]\,\mathrm{MHz}$ when randomly selecting the 500 detuning used in our dataset. The increase in noise for larger detunings is currently being investigated, but we do not think it affects the results we presented in the main text.
\label{apx:detunings}

\bibliography{BE_paper}

\begin{thebibliography}{40}%
\makeatletter
\providecommand \@ifxundefined [1]{%
 \@ifx{#1\undefined}
}%
\providecommand \@ifnum [1]{%
 \ifnum #1\expandafter \@firstoftwo
 \else \expandafter \@secondoftwo
 \fi
}%
\providecommand \@ifx [1]{%
 \ifx #1\expandafter \@firstoftwo
 \else \expandafter \@secondoftwo
 \fi
}%
\providecommand \natexlab [1]{#1}%
\providecommand \enquote  [1]{``#1''}%
\providecommand \bibnamefont  [1]{#1}%
\providecommand \bibfnamefont [1]{#1}%
\providecommand \citenamefont [1]{#1}%
\providecommand \href@noop [0]{\@secondoftwo}%
\providecommand \href [0]{\begingroup \@sanitize@url \@href}%
\providecommand \@href[1]{\@@startlink{#1}\@@href}%
\providecommand \@@href[1]{\endgroup#1\@@endlink}%
\providecommand \@sanitize@url [0]{\catcode `\\12\catcode `\$12\catcode
  `\&12\catcode `\#12\catcode `\^12\catcode `\_12\catcode `\%12\relax}%
\providecommand \@@startlink[1]{}%
\providecommand \@@endlink[0]{}%
\providecommand \url  [0]{\begingroup\@sanitize@url \@url }%
\providecommand \@url [1]{\endgroup\@href {#1}{\urlprefix }}%
\providecommand \urlprefix  [0]{URL }%
\providecommand \Eprint [0]{\href }%
\providecommand \doibase [0]{https://doi.org/}%
\providecommand \selectlanguage [0]{\@gobble}%
\providecommand \bibinfo  [0]{\@secondoftwo}%
\providecommand \bibfield  [0]{\@secondoftwo}%
\providecommand \translation [1]{[#1]}%
\providecommand \BibitemOpen [0]{}%
\providecommand \bibitemStop [0]{}%
\providecommand \bibitemNoStop [0]{.\EOS\space}%
\providecommand \EOS [0]{\spacefactor3000\relax}%
\providecommand \BibitemShut  [1]{\csname bibitem#1\endcsname}%
\let\auto@bib@innerbib\@empty
\bibitem [{\citenamefont {Crawford}\ \emph {et~al.}(2021)\citenamefont
  {Crawford}, \citenamefont {Shugayev}, \citenamefont {Paudel}, \citenamefont
  {Lu}, \citenamefont {Syamlal}, \citenamefont {Ohodnicki}, \citenamefont
  {Chorpening}, \citenamefont {Gentry},\ and\ \citenamefont
  {Duan}}]{Crawford2021}%
  \BibitemOpen
  \bibfield  {author} {\bibinfo {author} {\bibfnamefont {S.~E.}\ \bibnamefont
  {Crawford}}, \bibinfo {author} {\bibfnamefont {R.~A.}\ \bibnamefont
  {Shugayev}}, \bibinfo {author} {\bibfnamefont {H.~P.}\ \bibnamefont
  {Paudel}}, \bibinfo {author} {\bibfnamefont {P.}~\bibnamefont {Lu}}, \bibinfo
  {author} {\bibfnamefont {M.}~\bibnamefont {Syamlal}}, \bibinfo {author}
  {\bibfnamefont {P.~R.}\ \bibnamefont {Ohodnicki}}, \bibinfo {author}
  {\bibfnamefont {B.}~\bibnamefont {Chorpening}}, \bibinfo {author}
  {\bibfnamefont {R.}~\bibnamefont {Gentry}},\ and\ \bibinfo {author}
  {\bibfnamefont {Y.}~\bibnamefont {Duan}},\ }\bibfield  {title} {\bibinfo
  {title} {Quantum sensing for energy applications: Review and perspective},\
  }\href {https://doi.org/10.1002/qute.202100049} {\bibfield  {journal}
  {\bibinfo  {journal} {Adv. Quantum Technol.}\ }\textbf {\bibinfo {volume}
  {4}},\ \bibinfo {pages} {2100049} (\bibinfo {year} {2021})}\BibitemShut
  {NoStop}%
\bibitem [{\citenamefont {van~der Sar}\ \emph {et~al.}(2015)\citenamefont
  {van~der Sar}, \citenamefont {Casola}, \citenamefont {Walsworth},\ and\
  \citenamefont {Yacoby}}]{Sar2015}%
  \BibitemOpen
  \bibfield  {author} {\bibinfo {author} {\bibfnamefont {T.}~\bibnamefont
  {van~der Sar}}, \bibinfo {author} {\bibfnamefont {F.}~\bibnamefont {Casola}},
  \bibinfo {author} {\bibfnamefont {R.}~\bibnamefont {Walsworth}},\ and\
  \bibinfo {author} {\bibfnamefont {A.}~\bibnamefont {Yacoby}},\ }\bibfield
  {title} {\bibinfo {title} {Nanometre-scale probing of spin waves using single
  electron spins},\ }\href {https://doi.org/10.1038/ncomms8886} {\bibfield
  {journal} {\bibinfo  {journal} {Nat. Commun.}\ }\textbf {\bibinfo {volume}
  {6}},\ \bibinfo {pages} {7886} (\bibinfo {year} {2015})}\BibitemShut
  {NoStop}%
\bibitem [{\citenamefont {Gross}\ \emph {et~al.}(2017)\citenamefont {Gross},
  \citenamefont {Akhtar}, \citenamefont {Garcia}, \citenamefont {Martínez},
  \citenamefont {Chouaieb}, \citenamefont {Garcia}, \citenamefont
  {Carr\'et\'ero}, \citenamefont {Barth\'el\'emy}, \citenamefont {Appel},
  \citenamefont {Maletinsky}, \citenamefont {Kim}, \citenamefont {Chauleau},
  \citenamefont {Jaouen}, \citenamefont {Viret}, \citenamefont {Bibes},
  \citenamefont {Fusil},\ and\ \citenamefont {Jacques}}]{Gross2017}%
  \BibitemOpen
  \bibfield  {author} {\bibinfo {author} {\bibfnamefont {I.}~\bibnamefont
  {Gross}}, \bibinfo {author} {\bibfnamefont {W.}~\bibnamefont {Akhtar}},
  \bibinfo {author} {\bibfnamefont {V.}~\bibnamefont {Garcia}}, \bibinfo
  {author} {\bibfnamefont {L.~J.}\ \bibnamefont {Martínez}}, \bibinfo {author}
  {\bibfnamefont {S.}~\bibnamefont {Chouaieb}}, \bibinfo {author}
  {\bibfnamefont {K.}~\bibnamefont {Garcia}}, \bibinfo {author} {\bibfnamefont
  {C.}~\bibnamefont {Carr\'et\'ero}}, \bibinfo {author} {\bibfnamefont
  {A.}~\bibnamefont {Barth\'el\'emy}}, \bibinfo {author} {\bibfnamefont
  {P.}~\bibnamefont {Appel}}, \bibinfo {author} {\bibfnamefont
  {P.}~\bibnamefont {Maletinsky}}, \bibinfo {author} {\bibfnamefont {J.-V.}\
  \bibnamefont {Kim}}, \bibinfo {author} {\bibfnamefont {J.~Y.}\ \bibnamefont
  {Chauleau}}, \bibinfo {author} {\bibfnamefont {N.}~\bibnamefont {Jaouen}},
  \bibinfo {author} {\bibfnamefont {M.}~\bibnamefont {Viret}}, \bibinfo
  {author} {\bibfnamefont {M.}~\bibnamefont {Bibes}}, \bibinfo {author}
  {\bibfnamefont {S.}~\bibnamefont {Fusil}},\ and\ \bibinfo {author}
  {\bibfnamefont {V.}~\bibnamefont {Jacques}},\ }\bibfield  {title} {\bibinfo
  {title} {Real-space imaging of non-collinear antiferromagnetic order with a
  single-spin magnetometer},\ }\href {https://doi.org/10.1038/nature23656}
  {\bibfield  {journal} {\bibinfo  {journal} {Nature}\ }\textbf {\bibinfo
  {volume} {549}},\ \bibinfo {pages} {252} (\bibinfo {year}
  {2017})}\BibitemShut {NoStop}%
\bibitem [{\citenamefont {Dovzhenko}\ \emph {et~al.}(2018)\citenamefont
  {Dovzhenko}, \citenamefont {Casola}, \citenamefont {Schlotter}, \citenamefont
  {Zhou}, \citenamefont {B\"{u}ttner}, \citenamefont {Walsworth}, \citenamefont
  {Beach},\ and\ \citenamefont {Yacoby}}]{Dovzhenko2018}%
  \BibitemOpen
  \bibfield  {author} {\bibinfo {author} {\bibfnamefont {Y.}~\bibnamefont
  {Dovzhenko}}, \bibinfo {author} {\bibfnamefont {F.}~\bibnamefont {Casola}},
  \bibinfo {author} {\bibfnamefont {S.}~\bibnamefont {Schlotter}}, \bibinfo
  {author} {\bibfnamefont {T.~X.}\ \bibnamefont {Zhou}}, \bibinfo {author}
  {\bibfnamefont {F.}~\bibnamefont {B\"{u}ttner}}, \bibinfo {author}
  {\bibfnamefont {R.~L.}\ \bibnamefont {Walsworth}}, \bibinfo {author}
  {\bibfnamefont {G.~S.~D.}\ \bibnamefont {Beach}},\ and\ \bibinfo {author}
  {\bibfnamefont {A.}~\bibnamefont {Yacoby}},\ }\bibfield  {title} {\bibinfo
  {title} {Magnetostatic twists in room-temperature skyrmions explored by
  nitrogen-vacancy center spin texture reconstruction},\ }\href
  {https://doi.org/10.1038/s41467-018-05158-9} {\bibfield  {journal} {\bibinfo
  {journal} {Nat. Commun.}\ }\textbf {\bibinfo {volume} {9}},\ \bibinfo {pages}
  {2712} (\bibinfo {year} {2018})}\BibitemShut {NoStop}%
\bibitem [{\citenamefont {Jenkins}\ \emph {et~al.}(2019)\citenamefont
  {Jenkins}, \citenamefont {Pelliccione}, \citenamefont {Yu}, \citenamefont
  {Ma}, \citenamefont {Li}, \citenamefont {Wang},\ and\ \citenamefont
  {Jayich}}]{Jenkins2019}%
  \BibitemOpen
  \bibfield  {author} {\bibinfo {author} {\bibfnamefont {A.}~\bibnamefont
  {Jenkins}}, \bibinfo {author} {\bibfnamefont {M.}~\bibnamefont
  {Pelliccione}}, \bibinfo {author} {\bibfnamefont {G.}~\bibnamefont {Yu}},
  \bibinfo {author} {\bibfnamefont {X.}~\bibnamefont {Ma}}, \bibinfo {author}
  {\bibfnamefont {X.}~\bibnamefont {Li}}, \bibinfo {author} {\bibfnamefont
  {K.~L.}\ \bibnamefont {Wang}},\ and\ \bibinfo {author} {\bibfnamefont
  {A.~C.~B.}\ \bibnamefont {Jayich}},\ }\bibfield  {title} {\bibinfo {title}
  {Single-spin sensing of domain-wall structure and dynamics in a thin-film
  skyrmion host},\ }\href {https://doi.org/10.1103/physrevmaterials.3.083801}
  {\bibfield  {journal} {\bibinfo  {journal} {Phys. Rev. Mater.}\ }\textbf
  {\bibinfo {volume} {3}},\ \bibinfo {pages} {083801} (\bibinfo {year}
  {2019})}\BibitemShut {NoStop}%
\bibitem [{\citenamefont {Shi}\ \emph {et~al.}(2015)\citenamefont {Shi},
  \citenamefont {Zhang}, \citenamefont {Wang}, \citenamefont {Sun},
  \citenamefont {Wang}, \citenamefont {Rong}, \citenamefont {Chen},
  \citenamefont {Ju}, \citenamefont {Reinhard}, \citenamefont {Chen},
  \citenamefont {Wrachtrup}, \citenamefont {Wang},\ and\ \citenamefont
  {Du}}]{Shi2015}%
  \BibitemOpen
  \bibfield  {author} {\bibinfo {author} {\bibfnamefont {F.}~\bibnamefont
  {Shi}}, \bibinfo {author} {\bibfnamefont {Q.}~\bibnamefont {Zhang}}, \bibinfo
  {author} {\bibfnamefont {P.}~\bibnamefont {Wang}}, \bibinfo {author}
  {\bibfnamefont {H.}~\bibnamefont {Sun}}, \bibinfo {author} {\bibfnamefont
  {J.}~\bibnamefont {Wang}}, \bibinfo {author} {\bibfnamefont {X.}~\bibnamefont
  {Rong}}, \bibinfo {author} {\bibfnamefont {M.}~\bibnamefont {Chen}}, \bibinfo
  {author} {\bibfnamefont {C.}~\bibnamefont {Ju}}, \bibinfo {author}
  {\bibfnamefont {F.}~\bibnamefont {Reinhard}}, \bibinfo {author}
  {\bibfnamefont {H.}~\bibnamefont {Chen}}, \bibinfo {author} {\bibfnamefont
  {J.}~\bibnamefont {Wrachtrup}}, \bibinfo {author} {\bibfnamefont
  {J.}~\bibnamefont {Wang}},\ and\ \bibinfo {author} {\bibfnamefont
  {J.}~\bibnamefont {Du}},\ }\bibfield  {title} {\bibinfo {title}
  {Single-protein spin resonance spectroscopy under ambient conditions},\
  }\href {https://doi.org/10.1126/science.aaa2253} {\bibfield  {journal}
  {\bibinfo  {journal} {Science}\ }\textbf {\bibinfo {volume} {347}},\ \bibinfo
  {pages} {1135} (\bibinfo {year} {2015})}\BibitemShut {NoStop}%
\bibitem [{\citenamefont {Lovchinsky}\ \emph {et~al.}(2016)\citenamefont
  {Lovchinsky}, \citenamefont {Sushkov}, \citenamefont {Urbach}, \citenamefont
  {de~Leon}, \citenamefont {Choi}, \citenamefont {Greve}, \citenamefont
  {Evans}, \citenamefont {Gertner}, \citenamefont {Bersin}, \citenamefont
  {Müller}, \citenamefont {McGuinness}, \citenamefont {Jelezko}, \citenamefont
  {Walsworth}, \citenamefont {Park},\ and\ \citenamefont
  {Lukin}}]{Lovchinsky2016}%
  \BibitemOpen
  \bibfield  {author} {\bibinfo {author} {\bibfnamefont {I.}~\bibnamefont
  {Lovchinsky}}, \bibinfo {author} {\bibfnamefont {A.~O.}\ \bibnamefont
  {Sushkov}}, \bibinfo {author} {\bibfnamefont {E.}~\bibnamefont {Urbach}},
  \bibinfo {author} {\bibfnamefont {N.~P.}\ \bibnamefont {de~Leon}}, \bibinfo
  {author} {\bibfnamefont {S.}~\bibnamefont {Choi}}, \bibinfo {author}
  {\bibfnamefont {K.~D.}\ \bibnamefont {Greve}}, \bibinfo {author}
  {\bibfnamefont {R.}~\bibnamefont {Evans}}, \bibinfo {author} {\bibfnamefont
  {R.}~\bibnamefont {Gertner}}, \bibinfo {author} {\bibfnamefont
  {E.}~\bibnamefont {Bersin}}, \bibinfo {author} {\bibfnamefont
  {C.}~\bibnamefont {Müller}}, \bibinfo {author} {\bibfnamefont
  {L.}~\bibnamefont {McGuinness}}, \bibinfo {author} {\bibfnamefont
  {F.}~\bibnamefont {Jelezko}}, \bibinfo {author} {\bibfnamefont {R.~L.}\
  \bibnamefont {Walsworth}}, \bibinfo {author} {\bibfnamefont {H.}~\bibnamefont
  {Park}},\ and\ \bibinfo {author} {\bibfnamefont {M.~D.}\ \bibnamefont
  {Lukin}},\ }\bibfield  {title} {\bibinfo {title} {Nuclear magnetic resonance
  detection and spectroscopy of single proteins using quantum logic},\ }\href
  {https://doi.org/10.1126/science.aad8022} {\bibfield  {journal} {\bibinfo
  {journal} {Science}\ }\textbf {\bibinfo {volume} {351}},\ \bibinfo {pages}
  {836} (\bibinfo {year} {2016})}\BibitemShut {NoStop}%
\bibitem [{\citenamefont {Barry}\ \emph {et~al.}(2016)\citenamefont {Barry},
  \citenamefont {Turner}, \citenamefont {Schloss}, \citenamefont {Glenn},
  \citenamefont {Song}, \citenamefont {Lukin}, \citenamefont {Park},\ and\
  \citenamefont {Walsworth}}]{Barry2016}%
  \BibitemOpen
  \bibfield  {author} {\bibinfo {author} {\bibfnamefont {J.~F.}\ \bibnamefont
  {Barry}}, \bibinfo {author} {\bibfnamefont {M.~J.}\ \bibnamefont {Turner}},
  \bibinfo {author} {\bibfnamefont {J.~M.}\ \bibnamefont {Schloss}}, \bibinfo
  {author} {\bibfnamefont {D.~R.}\ \bibnamefont {Glenn}}, \bibinfo {author}
  {\bibfnamefont {Y.}~\bibnamefont {Song}}, \bibinfo {author} {\bibfnamefont
  {M.~D.}\ \bibnamefont {Lukin}}, \bibinfo {author} {\bibfnamefont
  {H.}~\bibnamefont {Park}},\ and\ \bibinfo {author} {\bibfnamefont {R.~L.}\
  \bibnamefont {Walsworth}},\ }\bibfield  {title} {\bibinfo {title} {Optical
  magnetic detection of single-neuron action potentials using quantum defects
  in diamond},\ }\href {https://doi.org/10.1073/pnas.1601513113} {\bibfield
  {journal} {\bibinfo  {journal} {Proc. Natl. Acad. Sci.}\ }\textbf {\bibinfo
  {volume} {113}},\ \bibinfo {pages} {14133} (\bibinfo {year}
  {2016})}\BibitemShut {NoStop}%
\bibitem [{\citenamefont {Sch\"afer-Nolte}\ \emph {et~al.}(2014)\citenamefont
  {Sch\"afer-Nolte}, \citenamefont {Schlipf}, \citenamefont {Ternes},
  \citenamefont {Reinhard}, \citenamefont {Kern},\ and\ \citenamefont
  {Wrachtrup}}]{SchaeferNolte2014}%
  \BibitemOpen
  \bibfield  {author} {\bibinfo {author} {\bibfnamefont {E.}~\bibnamefont
  {Sch\"afer-Nolte}}, \bibinfo {author} {\bibfnamefont {L.}~\bibnamefont
  {Schlipf}}, \bibinfo {author} {\bibfnamefont {M.}~\bibnamefont {Ternes}},
  \bibinfo {author} {\bibfnamefont {F.}~\bibnamefont {Reinhard}}, \bibinfo
  {author} {\bibfnamefont {K.}~\bibnamefont {Kern}},\ and\ \bibinfo {author}
  {\bibfnamefont {J.}~\bibnamefont {Wrachtrup}},\ }\bibfield  {title} {\bibinfo
  {title} {Tracking temperature-dependent relaxation times of ferritin
  nanomagnets with a wideband quantum spectrometer},\ }\href
  {https://doi.org/10.1103/physrevlett.113.217204} {\bibfield  {journal}
  {\bibinfo  {journal} {Phys. Rev. Lett.}\ }\textbf {\bibinfo {volume} {113}},\
  \bibinfo {pages} {217204} (\bibinfo {year} {2014})}\BibitemShut {NoStop}%
\bibitem [{\citenamefont {Finkler}\ and\ \citenamefont
  {Dasari}(2021)}]{Finkler2021}%
  \BibitemOpen
  \bibfield  {author} {\bibinfo {author} {\bibfnamefont {A.}~\bibnamefont
  {Finkler}}\ and\ \bibinfo {author} {\bibfnamefont {D.}~\bibnamefont
  {Dasari}},\ }\bibfield  {title} {\bibinfo {title} {Quantum sensing and
  control of spin-state dynamics in the radical-pair mechanism},\ }\href
  {https://doi.org/10.1103/physrevapplied.15.034066} {\bibfield  {journal}
  {\bibinfo  {journal} {Phys. Rev. Appl.}\ }\textbf {\bibinfo {volume} {15}},\
  \bibinfo {pages} {034066} (\bibinfo {year} {2021})}\BibitemShut {NoStop}%
\bibitem [{\citenamefont {Degen}\ \emph {et~al.}(2017)\citenamefont {Degen},
  \citenamefont {Reinhard},\ and\ \citenamefont {Cappellaro}}]{Degen2017}%
  \BibitemOpen
  \bibfield  {author} {\bibinfo {author} {\bibfnamefont {C.~L.}\ \bibnamefont
  {Degen}}, \bibinfo {author} {\bibfnamefont {F.}~\bibnamefont {Reinhard}},\
  and\ \bibinfo {author} {\bibfnamefont {P.}~\bibnamefont {Cappellaro}},\
  }\bibfield  {title} {\bibinfo {title} {Quantum sensing},\ }\href
  {https://doi.org/10.1103/RevModPhys.89.035002} {\bibfield  {journal}
  {\bibinfo  {journal} {Rev. Mod. Phys.}\ }\textbf {\bibinfo {volume} {89}},\
  \bibinfo {pages} {035002} (\bibinfo {year} {2017})}\BibitemShut {NoStop}%
\bibitem [{\citenamefont {Neumann}\ \emph {et~al.}(2013)\citenamefont
  {Neumann}, \citenamefont {Jakobi}, \citenamefont {Dolde}, \citenamefont
  {Burk}, \citenamefont {Reuter}, \citenamefont {Waldherr}, \citenamefont
  {Honert}, \citenamefont {Wolf}, \citenamefont {Brunner}, \citenamefont
  {Shim}, \citenamefont {Suter}, \citenamefont {Sumiya}, \citenamefont
  {Isoya},\ and\ \citenamefont {Wrachtrup}}]{Neumann2013}%
  \BibitemOpen
  \bibfield  {author} {\bibinfo {author} {\bibfnamefont {P.}~\bibnamefont
  {Neumann}}, \bibinfo {author} {\bibfnamefont {I.}~\bibnamefont {Jakobi}},
  \bibinfo {author} {\bibfnamefont {F.}~\bibnamefont {Dolde}}, \bibinfo
  {author} {\bibfnamefont {C.}~\bibnamefont {Burk}}, \bibinfo {author}
  {\bibfnamefont {R.}~\bibnamefont {Reuter}}, \bibinfo {author} {\bibfnamefont
  {G.}~\bibnamefont {Waldherr}}, \bibinfo {author} {\bibfnamefont
  {J.}~\bibnamefont {Honert}}, \bibinfo {author} {\bibfnamefont
  {T.}~\bibnamefont {Wolf}}, \bibinfo {author} {\bibfnamefont {A.}~\bibnamefont
  {Brunner}}, \bibinfo {author} {\bibfnamefont {J.~H.}\ \bibnamefont {Shim}},
  \bibinfo {author} {\bibfnamefont {D.}~\bibnamefont {Suter}}, \bibinfo
  {author} {\bibfnamefont {H.}~\bibnamefont {Sumiya}}, \bibinfo {author}
  {\bibfnamefont {J.}~\bibnamefont {Isoya}},\ and\ \bibinfo {author}
  {\bibfnamefont {J.}~\bibnamefont {Wrachtrup}},\ }\bibfield  {title} {\bibinfo
  {title} {High-precision nanoscale temperature sensing using single defects in
  diamond},\ }\href {https://doi.org/10.1021/nl401216y} {\bibfield  {journal}
  {\bibinfo  {journal} {Nano Lett.}\ }\textbf {\bibinfo {volume} {13}},\
  \bibinfo {pages} {2738} (\bibinfo {year} {2013})}\BibitemShut {NoStop}%
\bibitem [{\citenamefont {Trusheim}\ and\ \citenamefont
  {Englund}(2016)}]{Trusheim2016}%
  \BibitemOpen
  \bibfield  {author} {\bibinfo {author} {\bibfnamefont {M.~E.}\ \bibnamefont
  {Trusheim}}\ and\ \bibinfo {author} {\bibfnamefont {D.}~\bibnamefont
  {Englund}},\ }\bibfield  {title} {\bibinfo {title} {Wide-field strain imaging
  with preferentially-aligned nitrogen-vacancy centers in polycrystalline
  diamond},\ }\href {https://doi.org/10.1088/1367-2630/aa5040} {\bibfield
  {journal} {\bibinfo  {journal} {New J. Phys.}\ }\textbf {\bibinfo {volume}
  {18}},\ \bibinfo {pages} {123023} (\bibinfo {year} {2016})}\BibitemShut
  {NoStop}%
\bibitem [{\citenamefont {Ho}\ \emph {et~al.}(2021)\citenamefont {Ho},
  \citenamefont {Wong}, \citenamefont {Leung}, \citenamefont {Pang},
  \citenamefont {Leung}, \citenamefont {Yip}, \citenamefont {Zhang},
  \citenamefont {Xie}, \citenamefont {Goh},\ and\ \citenamefont
  {Yang}}]{Ho2021}%
  \BibitemOpen
  \bibfield  {author} {\bibinfo {author} {\bibfnamefont {K.~O.}\ \bibnamefont
  {Ho}}, \bibinfo {author} {\bibfnamefont {K.~C.}\ \bibnamefont {Wong}},
  \bibinfo {author} {\bibfnamefont {M.~Y.}\ \bibnamefont {Leung}}, \bibinfo
  {author} {\bibfnamefont {Y.~Y.}\ \bibnamefont {Pang}}, \bibinfo {author}
  {\bibfnamefont {W.~K.}\ \bibnamefont {Leung}}, \bibinfo {author}
  {\bibfnamefont {K.~Y.}\ \bibnamefont {Yip}}, \bibinfo {author} {\bibfnamefont
  {W.}~\bibnamefont {Zhang}}, \bibinfo {author} {\bibfnamefont
  {J.}~\bibnamefont {Xie}}, \bibinfo {author} {\bibfnamefont {S.~K.}\
  \bibnamefont {Goh}},\ and\ \bibinfo {author} {\bibfnamefont {S.}~\bibnamefont
  {Yang}},\ }\bibfield  {title} {\bibinfo {title} {Recent developments of
  quantum sensing under pressurized environment using the nitrogen vacancy
  ({NV}) center in diamond},\ }\href {https://doi.org/10.1063/5.0052233}
  {\bibfield  {journal} {\bibinfo  {journal} {J. Appl. Phys.}\ }\textbf
  {\bibinfo {volume} {129}},\ \bibinfo {pages} {241101} (\bibinfo {year}
  {2021})}\BibitemShut {NoStop}%
\bibitem [{\citenamefont {Higgins}\ \emph {et~al.}(2007)\citenamefont
  {Higgins}, \citenamefont {Berry}, \citenamefont {Bartlett}, \citenamefont
  {Wiseman},\ and\ \citenamefont {Pryde}}]{Higgins2007}%
  \BibitemOpen
  \bibfield  {author} {\bibinfo {author} {\bibfnamefont {B.~L.}\ \bibnamefont
  {Higgins}}, \bibinfo {author} {\bibfnamefont {D.~W.}\ \bibnamefont {Berry}},
  \bibinfo {author} {\bibfnamefont {S.~D.}\ \bibnamefont {Bartlett}}, \bibinfo
  {author} {\bibfnamefont {H.~M.}\ \bibnamefont {Wiseman}},\ and\ \bibinfo
  {author} {\bibfnamefont {G.~J.}\ \bibnamefont {Pryde}},\ }\bibfield  {title}
  {\bibinfo {title} {Entanglement-free {H}eisenberg-limited phase estimation},\
  }\href {https://doi.org/10.1038/nature06257} {\bibfield  {journal} {\bibinfo
  {journal} {Nature}\ }\textbf {\bibinfo {volume} {450}},\ \bibinfo {pages}
  {393} (\bibinfo {year} {2007})}\BibitemShut {NoStop}%
\bibitem [{\citenamefont {Bollinger}\ \emph {et~al.}(1996)\citenamefont
  {Bollinger}, \citenamefont {Itano}, \citenamefont {Wineland},\ and\
  \citenamefont {Heinzen}}]{Bollinger1996}%
  \BibitemOpen
  \bibfield  {author} {\bibinfo {author} {\bibfnamefont {J.~J.}\ \bibnamefont
  {Bollinger}}, \bibinfo {author} {\bibfnamefont {W.~M.}\ \bibnamefont
  {Itano}}, \bibinfo {author} {\bibfnamefont {D.~J.}\ \bibnamefont
  {Wineland}},\ and\ \bibinfo {author} {\bibfnamefont {D.~J.}\ \bibnamefont
  {Heinzen}},\ }\bibfield  {title} {\bibinfo {title} {Optimal frequency
  measurements with maximally correlated states},\ }\href
  {https://doi.org/10.1103/physreva.54.r4649} {\bibfield  {journal} {\bibinfo
  {journal} {Phys. Rev. A}\ }\textbf {\bibinfo {volume} {54}},\ \bibinfo
  {pages} {R4649} (\bibinfo {year} {1996})}\BibitemShut {NoStop}%
\bibitem [{\citenamefont {Vorobyov}\ \emph {et~al.}(2021)\citenamefont
  {Vorobyov}, \citenamefont {Zaiser}, \citenamefont {Abt}, \citenamefont
  {Meinel}, \citenamefont {Dasari}, \citenamefont {Neumann},\ and\
  \citenamefont {Wrachtrup}}]{Vorobyov2021}%
  \BibitemOpen
  \bibfield  {author} {\bibinfo {author} {\bibfnamefont {V.}~\bibnamefont
  {Vorobyov}}, \bibinfo {author} {\bibfnamefont {S.}~\bibnamefont {Zaiser}},
  \bibinfo {author} {\bibfnamefont {N.}~\bibnamefont {Abt}}, \bibinfo {author}
  {\bibfnamefont {J.}~\bibnamefont {Meinel}}, \bibinfo {author} {\bibfnamefont
  {D.}~\bibnamefont {Dasari}}, \bibinfo {author} {\bibfnamefont
  {P.}~\bibnamefont {Neumann}},\ and\ \bibinfo {author} {\bibfnamefont
  {J.}~\bibnamefont {Wrachtrup}},\ }\bibfield  {title} {\bibinfo {title}
  {Quantum {F}ourier transform for nanoscale quantum sensing},\ }\href
  {https://doi.org/10.1038/s41534-021-00463-6} {\bibfield  {journal} {\bibinfo
  {journal} {npj Quantum Inf.}\ }\textbf {\bibinfo {volume} {7}},\ \bibinfo
  {pages} {124} (\bibinfo {year} {2021})}\BibitemShut {NoStop}%
\bibitem [{\citenamefont {Kitaev}(1995)}]{Kitaev1995}%
  \BibitemOpen
  \bibfield  {author} {\bibinfo {author} {\bibfnamefont {A.~Y.}\ \bibnamefont
  {Kitaev}},\ }\bibfield  {title} {\bibinfo {title} {Quantum measurements and
  the {A}belian stabilizer problem},\ }\bibfield  {journal} {\bibinfo
  {journal} {arXiv}\ }\href {https://doi.org/10.48550/arXiv.quant-ph/9511026}
  {10.48550/arXiv.quant-ph/9511026} (\bibinfo {year} {1995})\BibitemShut
  {NoStop}%
\bibitem [{\citenamefont {Said}\ \emph {et~al.}(2011)\citenamefont {Said},
  \citenamefont {Berry},\ and\ \citenamefont {Twamley}}]{Said2011}%
  \BibitemOpen
  \bibfield  {author} {\bibinfo {author} {\bibfnamefont {R.~S.}\ \bibnamefont
  {Said}}, \bibinfo {author} {\bibfnamefont {D.~W.}\ \bibnamefont {Berry}},\
  and\ \bibinfo {author} {\bibfnamefont {J.}~\bibnamefont {Twamley}},\
  }\bibfield  {title} {\bibinfo {title} {Nanoscale magnetometry using a
  single-spin system in diamond},\ }\href
  {https://doi.org/10.1103/physrevb.83.125410} {\bibfield  {journal} {\bibinfo
  {journal} {Phys. Rev. B}\ }\textbf {\bibinfo {volume} {83}},\ \bibinfo
  {pages} {125410} (\bibinfo {year} {2011})}\BibitemShut {NoStop}%
\bibitem [{\citenamefont {Okamoto}\ \emph {et~al.}(2012)\citenamefont
  {Okamoto}, \citenamefont {Iefuji}, \citenamefont {Oyama}, \citenamefont
  {Yamagata}, \citenamefont {Imai}, \citenamefont {Fujiwara},\ and\
  \citenamefont {Takeuchi}}]{Okamoto2012}%
  \BibitemOpen
  \bibfield  {author} {\bibinfo {author} {\bibfnamefont {R.}~\bibnamefont
  {Okamoto}}, \bibinfo {author} {\bibfnamefont {M.}~\bibnamefont {Iefuji}},
  \bibinfo {author} {\bibfnamefont {S.}~\bibnamefont {Oyama}}, \bibinfo
  {author} {\bibfnamefont {K.}~\bibnamefont {Yamagata}}, \bibinfo {author}
  {\bibfnamefont {H.}~\bibnamefont {Imai}}, \bibinfo {author} {\bibfnamefont
  {A.}~\bibnamefont {Fujiwara}},\ and\ \bibinfo {author} {\bibfnamefont
  {S.}~\bibnamefont {Takeuchi}},\ }\bibfield  {title} {\bibinfo {title}
  {Experimental demonstration of adaptive quantum state estimation},\ }\href
  {https://doi.org/10.1103/physrevlett.109.130404} {\bibfield  {journal}
  {\bibinfo  {journal} {Phys. Rev. Lett.}\ }\textbf {\bibinfo {volume} {109}},\
  \bibinfo {pages} {130404} (\bibinfo {year} {2012})}\BibitemShut {NoStop}%
\bibitem [{\citenamefont {Danilin}\ \emph {et~al.}(2018)\citenamefont
  {Danilin}, \citenamefont {Lebedev}, \citenamefont {Vepsäläinen},
  \citenamefont {Lesovik}, \citenamefont {Blatter},\ and\ \citenamefont
  {Paraoanu}}]{Danilin2018}%
  \BibitemOpen
  \bibfield  {author} {\bibinfo {author} {\bibfnamefont {S.}~\bibnamefont
  {Danilin}}, \bibinfo {author} {\bibfnamefont {A.~V.}\ \bibnamefont
  {Lebedev}}, \bibinfo {author} {\bibfnamefont {A.}~\bibnamefont
  {Vepsäläinen}}, \bibinfo {author} {\bibfnamefont {G.~B.}\ \bibnamefont
  {Lesovik}}, \bibinfo {author} {\bibfnamefont {G.}~\bibnamefont {Blatter}},\
  and\ \bibinfo {author} {\bibfnamefont {G.~S.}\ \bibnamefont {Paraoanu}},\
  }\bibfield  {title} {\bibinfo {title} {Quantum-enhanced magnetometry by phase
  estimation algorithms with a single artificial atom},\ }\href
  {https://doi.org/10.1038/s41534-018-0078-y} {\bibfield  {journal} {\bibinfo
  {journal} {npj Quantum Inf.}\ }\textbf {\bibinfo {volume} {4}},\ \bibinfo
  {pages} {29} (\bibinfo {year} {2018})}\BibitemShut {NoStop}%
\bibitem [{\citenamefont {Kitaev}\ \emph {et~al.}(2002)\citenamefont {Kitaev},
  \citenamefont {Shen},\ and\ \citenamefont {Vyalyi}}]{Kitaev2002}%
  \BibitemOpen
  \bibfield  {author} {\bibinfo {author} {\bibfnamefont {A.}~\bibnamefont
  {Kitaev}}, \bibinfo {author} {\bibfnamefont {A.}~\bibnamefont {Shen}},\ and\
  \bibinfo {author} {\bibfnamefont {M.}~\bibnamefont {Vyalyi}},\ }\href
  {https://doi.org/10.1090/gsm/047} {\emph {\bibinfo {title} {Classical and
  Quantum Computation}}}\ (\bibinfo  {publisher} {American Mathematical
  Society},\ \bibinfo {year} {2002})\BibitemShut {NoStop}%
\bibitem [{\citenamefont {Bonato}\ \emph {et~al.}(2015)\citenamefont {Bonato},
  \citenamefont {Blok}, \citenamefont {Dinani}, \citenamefont {Berry},
  \citenamefont {Markham}, \citenamefont {Twitchen},\ and\ \citenamefont
  {Hanson}}]{Bonato2015}%
  \BibitemOpen
  \bibfield  {author} {\bibinfo {author} {\bibfnamefont {C.}~\bibnamefont
  {Bonato}}, \bibinfo {author} {\bibfnamefont {M.~S.}\ \bibnamefont {Blok}},
  \bibinfo {author} {\bibfnamefont {H.~T.}\ \bibnamefont {Dinani}}, \bibinfo
  {author} {\bibfnamefont {D.~W.}\ \bibnamefont {Berry}}, \bibinfo {author}
  {\bibfnamefont {M.~L.}\ \bibnamefont {Markham}}, \bibinfo {author}
  {\bibfnamefont {D.~J.}\ \bibnamefont {Twitchen}},\ and\ \bibinfo {author}
  {\bibfnamefont {R.}~\bibnamefont {Hanson}},\ }\bibfield  {title} {\bibinfo
  {title} {Optimized quantum sensing with a single electron spin using
  real-time adaptive measurements},\ }\href
  {https://doi.org/10.1038/nnano.2015.261} {\bibfield  {journal} {\bibinfo
  {journal} {Nat. Nanotechnol.}\ }\textbf {\bibinfo {volume} {11}},\ \bibinfo
  {pages} {247} (\bibinfo {year} {2015})}\BibitemShut {NoStop}%
\bibitem [{\citenamefont {Valeri}\ \emph {et~al.}(2020)\citenamefont {Valeri},
  \citenamefont {Polino}, \citenamefont {Poderini}, \citenamefont {Gianani},
  \citenamefont {Corrielli}, \citenamefont {Crespi}, \citenamefont {Osellame},
  \citenamefont {Spagnolo},\ and\ \citenamefont {Sciarrino}}]{Valeri2020}%
  \BibitemOpen
  \bibfield  {author} {\bibinfo {author} {\bibfnamefont {M.}~\bibnamefont
  {Valeri}}, \bibinfo {author} {\bibfnamefont {E.}~\bibnamefont {Polino}},
  \bibinfo {author} {\bibfnamefont {D.}~\bibnamefont {Poderini}}, \bibinfo
  {author} {\bibfnamefont {I.}~\bibnamefont {Gianani}}, \bibinfo {author}
  {\bibfnamefont {G.}~\bibnamefont {Corrielli}}, \bibinfo {author}
  {\bibfnamefont {A.}~\bibnamefont {Crespi}}, \bibinfo {author} {\bibfnamefont
  {R.}~\bibnamefont {Osellame}}, \bibinfo {author} {\bibfnamefont
  {N.}~\bibnamefont {Spagnolo}},\ and\ \bibinfo {author} {\bibfnamefont
  {F.}~\bibnamefont {Sciarrino}},\ }\bibfield  {title} {\bibinfo {title}
  {Experimental adaptive {B}ayesian estimation of multiple phases with limited
  data},\ }\href {https://doi.org/10.1038/s41534-020-00326-6} {\bibfield
  {journal} {\bibinfo  {journal} {npj Quantum Inf.}\ }\textbf {\bibinfo
  {volume} {6}},\ \bibinfo {pages} {92} (\bibinfo {year} {2020})}\BibitemShut
  {NoStop}%
\bibitem [{\citenamefont {Gebhart}\ \emph {et~al.}(2022)\citenamefont
  {Gebhart}, \citenamefont {Santagati}, \citenamefont {Gentile}, \citenamefont
  {Gauger}, \citenamefont {Craig}, \citenamefont {Ares}, \citenamefont
  {Banchi}, \citenamefont {Marquardt}, \citenamefont {Pezz\'e},\ and\
  \citenamefont {Bonato}}]{Gebhart2022}%
  \BibitemOpen
  \bibfield  {author} {\bibinfo {author} {\bibfnamefont {V.}~\bibnamefont
  {Gebhart}}, \bibinfo {author} {\bibfnamefont {R.}~\bibnamefont {Santagati}},
  \bibinfo {author} {\bibfnamefont {A.~A.}\ \bibnamefont {Gentile}}, \bibinfo
  {author} {\bibfnamefont {E.}~\bibnamefont {Gauger}}, \bibinfo {author}
  {\bibfnamefont {D.}~\bibnamefont {Craig}}, \bibinfo {author} {\bibfnamefont
  {N.}~\bibnamefont {Ares}}, \bibinfo {author} {\bibfnamefont {L.}~\bibnamefont
  {Banchi}}, \bibinfo {author} {\bibfnamefont {F.}~\bibnamefont {Marquardt}},
  \bibinfo {author} {\bibfnamefont {L.}~\bibnamefont {Pezz\'e}},\ and\ \bibinfo
  {author} {\bibfnamefont {C.}~\bibnamefont {Bonato}},\ }\bibfield  {title}
  {\bibinfo {title} {Learning quantum systems},\ }\Eprint
  {https://arxiv.org/abs/2207.00298} {arXiv:2207.00298}  (\bibinfo {year}
  {2022}),\ \bibinfo {note} {deposited on the arXiv.}\BibitemShut {Stop}%
\bibitem [{\citenamefont {Higgins}\ \emph {et~al.}(2009)\citenamefont
  {Higgins}, \citenamefont {Berry}, \citenamefont {Bartlett}, \citenamefont
  {Mitchell}, \citenamefont {Wiseman},\ and\ \citenamefont
  {Pryde}}]{Higgins2009}%
  \BibitemOpen
  \bibfield  {author} {\bibinfo {author} {\bibfnamefont {B.~L.}\ \bibnamefont
  {Higgins}}, \bibinfo {author} {\bibfnamefont {D.~W.}\ \bibnamefont {Berry}},
  \bibinfo {author} {\bibfnamefont {S.~D.}\ \bibnamefont {Bartlett}}, \bibinfo
  {author} {\bibfnamefont {M.~W.}\ \bibnamefont {Mitchell}}, \bibinfo {author}
  {\bibfnamefont {H.~M.}\ \bibnamefont {Wiseman}},\ and\ \bibinfo {author}
  {\bibfnamefont {G.~J.}\ \bibnamefont {Pryde}},\ }\bibfield  {title} {\bibinfo
  {title} {Demonstrating {H}eisenberg-limited unambiguous phase estimation
  without adaptive measurements},\ }\href
  {https://doi.org/10.1088/1367-2630/11/7/073023} {\bibfield  {journal}
  {\bibinfo  {journal} {New J. Phys.}\ }\textbf {\bibinfo {volume} {11}},\
  \bibinfo {pages} {073023} (\bibinfo {year} {2009})}\BibitemShut {NoStop}%
\bibitem [{\citenamefont {Cappellaro}(2012)}]{Cappellaro2012}%
  \BibitemOpen
  \bibfield  {author} {\bibinfo {author} {\bibfnamefont {P.}~\bibnamefont
  {Cappellaro}},\ }\bibfield  {title} {\bibinfo {title} {Spin-bath narrowing
  with adaptive parameter estimation},\ }\href
  {https://doi.org/10.1103/physreva.85.030301} {\bibfield  {journal} {\bibinfo
  {journal} {Phys. Rev. A}\ }\textbf {\bibinfo {volume} {85}},\ \bibinfo
  {pages} {030301} (\bibinfo {year} {2012})}\BibitemShut {NoStop}%
\bibitem [{\citenamefont {Wiebe}\ and\ \citenamefont
  {Granade}(2016)}]{Wiebe2016}%
  \BibitemOpen
  \bibfield  {author} {\bibinfo {author} {\bibfnamefont {N.}~\bibnamefont
  {Wiebe}}\ and\ \bibinfo {author} {\bibfnamefont {C.}~\bibnamefont
  {Granade}},\ }\bibfield  {title} {\bibinfo {title} {Efficient {B}ayesian
  phase estimation},\ }\href {https://doi.org/10.1103/physrevlett.117.010503}
  {\bibfield  {journal} {\bibinfo  {journal} {Phys. Rev. Lett.}\ }\textbf
  {\bibinfo {volume} {117}},\ \bibinfo {pages} {010503} (\bibinfo {year}
  {2016})}\BibitemShut {NoStop}%
\bibitem [{\citenamefont {Scerri}\ \emph {et~al.}(2020)\citenamefont {Scerri},
  \citenamefont {Gauger},\ and\ \citenamefont {Bonato}}]{Scerri2020}%
  \BibitemOpen
  \bibfield  {author} {\bibinfo {author} {\bibfnamefont {E.}~\bibnamefont
  {Scerri}}, \bibinfo {author} {\bibfnamefont {E.~M.}\ \bibnamefont {Gauger}},\
  and\ \bibinfo {author} {\bibfnamefont {C.}~\bibnamefont {Bonato}},\
  }\bibfield  {title} {\bibinfo {title} {Extending qubit coherence by adaptive
  quantum environment learning},\ }\href
  {https://doi.org/10.1088/1367-2630/ab7bf3} {\bibfield  {journal} {\bibinfo
  {journal} {New J. Phys.}\ }\textbf {\bibinfo {volume} {22}},\ \bibinfo
  {pages} {035002} (\bibinfo {year} {2020})}\BibitemShut {NoStop}%
\bibitem [{\citenamefont {Santagati}\ \emph {et~al.}(2019)\citenamefont
  {Santagati}, \citenamefont {Gentile}, \citenamefont {Knauer}, \citenamefont
  {Schmitt}, \citenamefont {Paesani}, \citenamefont {Granade}, \citenamefont
  {Wiebe}, \citenamefont {Osterkamp}, \citenamefont {McGuinness}, \citenamefont
  {Wang}, \citenamefont {Thompson}, \citenamefont {Rarity}, \citenamefont
  {Jelezko},\ and\ \citenamefont {Laing}}]{Santagati2019}%
  \BibitemOpen
  \bibfield  {author} {\bibinfo {author} {\bibfnamefont {R.}~\bibnamefont
  {Santagati}}, \bibinfo {author} {\bibfnamefont {A.}~\bibnamefont {Gentile}},
  \bibinfo {author} {\bibfnamefont {S.}~\bibnamefont {Knauer}}, \bibinfo
  {author} {\bibfnamefont {S.}~\bibnamefont {Schmitt}}, \bibinfo {author}
  {\bibfnamefont {S.}~\bibnamefont {Paesani}}, \bibinfo {author} {\bibfnamefont
  {C.}~\bibnamefont {Granade}}, \bibinfo {author} {\bibfnamefont
  {N.}~\bibnamefont {Wiebe}}, \bibinfo {author} {\bibfnamefont
  {C.}~\bibnamefont {Osterkamp}}, \bibinfo {author} {\bibfnamefont
  {L.}~\bibnamefont {McGuinness}}, \bibinfo {author} {\bibfnamefont
  {J.}~\bibnamefont {Wang}}, \bibinfo {author} {\bibfnamefont {M.}~\bibnamefont
  {Thompson}}, \bibinfo {author} {\bibfnamefont {J.}~\bibnamefont {Rarity}},
  \bibinfo {author} {\bibfnamefont {F.}~\bibnamefont {Jelezko}},\ and\ \bibinfo
  {author} {\bibfnamefont {A.}~\bibnamefont {Laing}},\ }\bibfield  {title}
  {\bibinfo {title} {Magnetic-field learning using a single electronic spin in
  diamond with one-photon readout at room temperature},\ }\href
  {https://doi.org/10.1103/physrevx.9.021019} {\bibfield  {journal} {\bibinfo
  {journal} {Phys. Rev. X}\ }\textbf {\bibinfo {volume} {9}},\ \bibinfo {pages}
  {021019} (\bibinfo {year} {2019})}\BibitemShut {NoStop}%
\bibitem [{\citenamefont {Joas}\ \emph {et~al.}(2021)\citenamefont {Joas},
  \citenamefont {Schmitt}, \citenamefont {Santagati}, \citenamefont {Gentile},
  \citenamefont {Bonato}, \citenamefont {Laing}, \citenamefont {McGuinness},\
  and\ \citenamefont {Jelezko}}]{Joas2021}%
  \BibitemOpen
  \bibfield  {author} {\bibinfo {author} {\bibfnamefont {T.}~\bibnamefont
  {Joas}}, \bibinfo {author} {\bibfnamefont {S.}~\bibnamefont {Schmitt}},
  \bibinfo {author} {\bibfnamefont {R.}~\bibnamefont {Santagati}}, \bibinfo
  {author} {\bibfnamefont {A.~A.}\ \bibnamefont {Gentile}}, \bibinfo {author}
  {\bibfnamefont {C.}~\bibnamefont {Bonato}}, \bibinfo {author} {\bibfnamefont
  {A.}~\bibnamefont {Laing}}, \bibinfo {author} {\bibfnamefont {L.~P.}\
  \bibnamefont {McGuinness}},\ and\ \bibinfo {author} {\bibfnamefont
  {F.}~\bibnamefont {Jelezko}},\ }\bibfield  {title} {\bibinfo {title} {Online
  adaptive quantum characterization of a nuclear spin},\ }\href
  {https://doi.org/10.1038/s41534-021-00389-z} {\bibfield  {journal} {\bibinfo
  {journal} {npj Quantum Inf.}\ }\textbf {\bibinfo {volume} {7}},\ \bibinfo
  {pages} {56} (\bibinfo {year} {2021})}\BibitemShut {NoStop}%
\bibitem [{\citenamefont {Dinani}\ \emph {et~al.}(2019)\citenamefont {Dinani},
  \citenamefont {Berry}, \citenamefont {Gonzalez}, \citenamefont {Maze},\ and\
  \citenamefont {Bonato}}]{Dinani2019}%
  \BibitemOpen
  \bibfield  {author} {\bibinfo {author} {\bibfnamefont {H.~T.}\ \bibnamefont
  {Dinani}}, \bibinfo {author} {\bibfnamefont {D.~W.}\ \bibnamefont {Berry}},
  \bibinfo {author} {\bibfnamefont {R.}~\bibnamefont {Gonzalez}}, \bibinfo
  {author} {\bibfnamefont {J.~R.}\ \bibnamefont {Maze}},\ and\ \bibinfo
  {author} {\bibfnamefont {C.}~\bibnamefont {Bonato}},\ }\bibfield  {title}
  {\bibinfo {title} {Bayesian estimation for quantum sensing in the absence of
  single-shot detection},\ }\href {https://doi.org/10.1103/physrevb.99.125413}
  {\bibfield  {journal} {\bibinfo  {journal} {Phys. Rev. B}\ }\textbf {\bibinfo
  {volume} {99}},\ \bibinfo {pages} {125413} (\bibinfo {year}
  {2019})}\BibitemShut {NoStop}%
\bibitem [{\citenamefont {Maze}\ \emph {et~al.}(2008)\citenamefont {Maze},
  \citenamefont {Stanwix}, \citenamefont {Hodges}, \citenamefont {Hong},
  \citenamefont {Taylor}, \citenamefont {Cappellaro}, \citenamefont {Jiang},
  \citenamefont {Dutt}, \citenamefont {Togan}, \citenamefont {Zibrov},
  \citenamefont {Yacoby}, \citenamefont {Walsworth},\ and\ \citenamefont
  {Lukin}}]{Maze2008}%
  \BibitemOpen
  \bibfield  {author} {\bibinfo {author} {\bibfnamefont {J.~R.}\ \bibnamefont
  {Maze}}, \bibinfo {author} {\bibfnamefont {P.~L.}\ \bibnamefont {Stanwix}},
  \bibinfo {author} {\bibfnamefont {J.~S.}\ \bibnamefont {Hodges}}, \bibinfo
  {author} {\bibfnamefont {S.}~\bibnamefont {Hong}}, \bibinfo {author}
  {\bibfnamefont {J.~M.}\ \bibnamefont {Taylor}}, \bibinfo {author}
  {\bibfnamefont {P.}~\bibnamefont {Cappellaro}}, \bibinfo {author}
  {\bibfnamefont {L.}~\bibnamefont {Jiang}}, \bibinfo {author} {\bibfnamefont
  {M.~V.~G.}\ \bibnamefont {Dutt}}, \bibinfo {author} {\bibfnamefont
  {E.}~\bibnamefont {Togan}}, \bibinfo {author} {\bibfnamefont {A.~S.}\
  \bibnamefont {Zibrov}}, \bibinfo {author} {\bibfnamefont {A.}~\bibnamefont
  {Yacoby}}, \bibinfo {author} {\bibfnamefont {R.~L.}\ \bibnamefont
  {Walsworth}},\ and\ \bibinfo {author} {\bibfnamefont {M.~D.}\ \bibnamefont
  {Lukin}},\ }\bibfield  {title} {\bibinfo {title} {Nanoscale magnetic sensing
  with an individual electronic spin in diamond},\ }\href
  {https://doi.org/10.1038/nature07279} {\bibfield  {journal} {\bibinfo
  {journal} {Nature}\ }\textbf {\bibinfo {volume} {455}},\ \bibinfo {pages}
  {644} (\bibinfo {year} {2008})}\BibitemShut {NoStop}%
\bibitem [{\citenamefont {Balasubramanian}\ \emph {et~al.}(2008)\citenamefont
  {Balasubramanian}, \citenamefont {Chan}, \citenamefont {Kolesov},
  \citenamefont {Al-Hmoud}, \citenamefont {Tisler}, \citenamefont {Shin},
  \citenamefont {Kim}, \citenamefont {Wojcik}, \citenamefont {Hemmer},
  \citenamefont {Krueger}, \citenamefont {Hanke}, \citenamefont
  {Leitenstorfer}, \citenamefont {Bratschitsch}, \citenamefont {Jelezko},\ and\
  \citenamefont {Wrachtrup}}]{Balasubramanian2008}%
  \BibitemOpen
  \bibfield  {author} {\bibinfo {author} {\bibfnamefont {G.}~\bibnamefont
  {Balasubramanian}}, \bibinfo {author} {\bibfnamefont {I.~Y.}\ \bibnamefont
  {Chan}}, \bibinfo {author} {\bibfnamefont {R.}~\bibnamefont {Kolesov}},
  \bibinfo {author} {\bibfnamefont {M.}~\bibnamefont {Al-Hmoud}}, \bibinfo
  {author} {\bibfnamefont {J.}~\bibnamefont {Tisler}}, \bibinfo {author}
  {\bibfnamefont {C.}~\bibnamefont {Shin}}, \bibinfo {author} {\bibfnamefont
  {C.}~\bibnamefont {Kim}}, \bibinfo {author} {\bibfnamefont {A.}~\bibnamefont
  {Wojcik}}, \bibinfo {author} {\bibfnamefont {P.~R.}\ \bibnamefont {Hemmer}},
  \bibinfo {author} {\bibfnamefont {A.}~\bibnamefont {Krueger}}, \bibinfo
  {author} {\bibfnamefont {T.}~\bibnamefont {Hanke}}, \bibinfo {author}
  {\bibfnamefont {A.}~\bibnamefont {Leitenstorfer}}, \bibinfo {author}
  {\bibfnamefont {R.}~\bibnamefont {Bratschitsch}}, \bibinfo {author}
  {\bibfnamefont {F.}~\bibnamefont {Jelezko}},\ and\ \bibinfo {author}
  {\bibfnamefont {J.}~\bibnamefont {Wrachtrup}},\ }\bibfield  {title} {\bibinfo
  {title} {Nanoscale imaging magnetometry with diamond spins under ambient
  conditions},\ }\href {https://doi.org/10.1038/nature07278} {\bibfield
  {journal} {\bibinfo  {journal} {Nature}\ }\textbf {\bibinfo {volume} {455}},\
  \bibinfo {pages} {648} (\bibinfo {year} {2008})}\BibitemShut {NoStop}%
\bibitem [{\citenamefont {Childress}\ \emph {et~al.}(2006)\citenamefont
  {Childress}, \citenamefont {Gurudev~Dutt}, \citenamefont {Taylor},
  \citenamefont {Zibrov}, \citenamefont {Jelezko}, \citenamefont {Wrachtrup},
  \citenamefont {Hemmer},\ and\ \citenamefont {Lukin}}]{Childress2006}%
  \BibitemOpen
  \bibfield  {author} {\bibinfo {author} {\bibfnamefont {L.}~\bibnamefont
  {Childress}}, \bibinfo {author} {\bibfnamefont {M.~V.}\ \bibnamefont
  {Gurudev~Dutt}}, \bibinfo {author} {\bibfnamefont {J.~M.}\ \bibnamefont
  {Taylor}}, \bibinfo {author} {\bibfnamefont {A.~S.}\ \bibnamefont {Zibrov}},
  \bibinfo {author} {\bibfnamefont {F.}~\bibnamefont {Jelezko}}, \bibinfo
  {author} {\bibfnamefont {J.}~\bibnamefont {Wrachtrup}}, \bibinfo {author}
  {\bibfnamefont {P.~R.}\ \bibnamefont {Hemmer}},\ and\ \bibinfo {author}
  {\bibfnamefont {M.~D.}\ \bibnamefont {Lukin}},\ }\bibfield  {title} {\bibinfo
  {title} {Coherent dynamics of coupled electron and nuclear spin qubits in
  diamond},\ }\href {https://doi.org/10.1126/science.1131871} {\bibfield
  {journal} {\bibinfo  {journal} {Science}\ }\textbf {\bibinfo {volume}
  {314}},\ \bibinfo {pages} {281} (\bibinfo {year} {2006})}\BibitemShut
  {NoStop}%
\bibitem [{\citenamefont {Robledo}\ \emph {et~al.}(2011)\citenamefont
  {Robledo}, \citenamefont {Childress}, \citenamefont {Bernien}, \citenamefont
  {Hensen}, \citenamefont {Alkemade},\ and\ \citenamefont
  {Hanson}}]{Robledo2011}%
  \BibitemOpen
  \bibfield  {author} {\bibinfo {author} {\bibfnamefont {L.}~\bibnamefont
  {Robledo}}, \bibinfo {author} {\bibfnamefont {L.}~\bibnamefont {Childress}},
  \bibinfo {author} {\bibfnamefont {H.}~\bibnamefont {Bernien}}, \bibinfo
  {author} {\bibfnamefont {B.}~\bibnamefont {Hensen}}, \bibinfo {author}
  {\bibfnamefont {P.~F.~A.}\ \bibnamefont {Alkemade}},\ and\ \bibinfo {author}
  {\bibfnamefont {R.}~\bibnamefont {Hanson}},\ }\bibfield  {title} {\bibinfo
  {title} {High-fidelity projective read-out of a solid-state spin quantum
  register},\ }\href {https://doi.org/10.1038/nature10401} {\bibfield
  {journal} {\bibinfo  {journal} {Nature}\ }\textbf {\bibinfo {volume} {477}},\
  \bibinfo {pages} {574} (\bibinfo {year} {2011})}\BibitemShut {NoStop}%
\bibitem [{\citenamefont {Staudacher}\ \emph {et~al.}(2013)\citenamefont
  {Staudacher}, \citenamefont {Shi}, \citenamefont {Pezzagna}, \citenamefont
  {Meijer}, \citenamefont {Du}, \citenamefont {Meriles}, \citenamefont
  {Reinhard},\ and\ \citenamefont {Wrachtrup}}]{Staudacher2013}%
  \BibitemOpen
  \bibfield  {author} {\bibinfo {author} {\bibfnamefont {T.}~\bibnamefont
  {Staudacher}}, \bibinfo {author} {\bibfnamefont {F.}~\bibnamefont {Shi}},
  \bibinfo {author} {\bibfnamefont {S.}~\bibnamefont {Pezzagna}}, \bibinfo
  {author} {\bibfnamefont {J.}~\bibnamefont {Meijer}}, \bibinfo {author}
  {\bibfnamefont {J.}~\bibnamefont {Du}}, \bibinfo {author} {\bibfnamefont
  {C.~A.}\ \bibnamefont {Meriles}}, \bibinfo {author} {\bibfnamefont
  {F.}~\bibnamefont {Reinhard}},\ and\ \bibinfo {author} {\bibfnamefont
  {J.}~\bibnamefont {Wrachtrup}},\ }\bibfield  {title} {\bibinfo {title}
  {Nuclear magnetic resonance spectroscopy on a (5-nanometer)$^3$ sample
  volume},\ }\href {https://doi.org/10.1126/science.1231675} {\bibfield
  {journal} {\bibinfo  {journal} {Science}\ }\textbf {\bibinfo {volume}
  {339}},\ \bibinfo {pages} {561} (\bibinfo {year} {2013})}\BibitemShut
  {NoStop}%
\bibitem [{\citenamefont {Momenzadeh}\ \emph {et~al.}(2015)\citenamefont
  {Momenzadeh}, \citenamefont {St\"{o}hr}, \citenamefont {de~Oliveira},
  \citenamefont {Brunner}, \citenamefont {Denisenko}, \citenamefont {Yang},
  \citenamefont {Reinhard},\ and\ \citenamefont {Wrachtrup}}]{Momenzadeh2015}%
  \BibitemOpen
  \bibfield  {author} {\bibinfo {author} {\bibfnamefont {S.~A.}\ \bibnamefont
  {Momenzadeh}}, \bibinfo {author} {\bibfnamefont {R.~J.}\ \bibnamefont
  {St\"{o}hr}}, \bibinfo {author} {\bibfnamefont {F.~F.}\ \bibnamefont
  {de~Oliveira}}, \bibinfo {author} {\bibfnamefont {A.}~\bibnamefont
  {Brunner}}, \bibinfo {author} {\bibfnamefont {A.}~\bibnamefont {Denisenko}},
  \bibinfo {author} {\bibfnamefont {S.}~\bibnamefont {Yang}}, \bibinfo {author}
  {\bibfnamefont {F.}~\bibnamefont {Reinhard}},\ and\ \bibinfo {author}
  {\bibfnamefont {J.}~\bibnamefont {Wrachtrup}},\ }\bibfield  {title} {\bibinfo
  {title} {Nanoengineered diamond waveguide as a robust bright platform for
  nanomagnetometry using shallow nitrogen vacancy centers},\ }\href
  {https://doi.org/10.1021/nl503326t} {\bibfield  {journal} {\bibinfo
  {journal} {Nano Lett.}\ }\textbf {\bibinfo {volume} {15}},\ \bibinfo {pages}
  {165} (\bibinfo {year} {2015})}\BibitemShut {NoStop}%
\bibitem [{\citenamefont {McMichael}\ \emph {et~al.}(2021)\citenamefont
  {McMichael}, \citenamefont {Dushenko},\ and\ \citenamefont
  {Blakley}}]{McMichael2021}%
  \BibitemOpen
  \bibfield  {author} {\bibinfo {author} {\bibfnamefont {R.~D.}\ \bibnamefont
  {McMichael}}, \bibinfo {author} {\bibfnamefont {S.}~\bibnamefont
  {Dushenko}},\ and\ \bibinfo {author} {\bibfnamefont {S.~M.}\ \bibnamefont
  {Blakley}},\ }\bibfield  {title} {\bibinfo {title} {Sequential {B}ayesian
  experiment design for adaptive {R}amsey sequence measurements},\ }\href
  {https://doi.org/10.1063/5.0055630} {\bibfield  {journal} {\bibinfo
  {journal} {J. Appl. Phys.}\ }\textbf {\bibinfo {volume} {130}},\ \bibinfo
  {pages} {144401} (\bibinfo {year} {2021})}\BibitemShut {NoStop}%
\bibitem [{\citenamefont {Arshad}\ \emph {et~al.}()\citenamefont {Arshad},
  \citenamefont {Bekker}, \citenamefont {Haylock}, \citenamefont {Skrzypczak},
  \citenamefont {White}, \citenamefont {Griffiths}, \citenamefont {Gore},
  \citenamefont {Morley}, \citenamefont {Salter}, \citenamefont {Smith},
  \citenamefont {Zohar}, \citenamefont {Finkler}, \citenamefont {Altmann},
  \citenamefont {Gauger},\ and\ \citenamefont {Bonato}}]{Arshad2022}%
  \BibitemOpen
  \bibfield  {author} {\bibinfo {author} {\bibfnamefont {M.~J.}\ \bibnamefont
  {Arshad}}, \bibinfo {author} {\bibfnamefont {C.}~\bibnamefont {Bekker}},
  \bibinfo {author} {\bibfnamefont {B.}~\bibnamefont {Haylock}}, \bibinfo
  {author} {\bibfnamefont {K.}~\bibnamefont {Skrzypczak}}, \bibinfo {author}
  {\bibfnamefont {D.}~\bibnamefont {White}}, \bibinfo {author} {\bibfnamefont
  {B.}~\bibnamefont {Griffiths}}, \bibinfo {author} {\bibfnamefont
  {J.}~\bibnamefont {Gore}}, \bibinfo {author} {\bibfnamefont {G.~W.}\
  \bibnamefont {Morley}}, \bibinfo {author} {\bibfnamefont {P.}~\bibnamefont
  {Salter}}, \bibinfo {author} {\bibfnamefont {J.}~\bibnamefont {Smith}},
  \bibinfo {author} {\bibfnamefont {I.}~\bibnamefont {Zohar}}, \bibinfo
  {author} {\bibfnamefont {A.}~\bibnamefont {Finkler}}, \bibinfo {author}
  {\bibfnamefont {Y.}~\bibnamefont {Altmann}}, \bibinfo {author} {\bibfnamefont
  {E.~M.}\ \bibnamefont {Gauger}},\ and\ \bibinfo {author} {\bibfnamefont
  {C.}~\bibnamefont {Bonato}},\ }\bibfield  {title} {\bibinfo {title} {Online
  adaptive estimation of decoherence timescales for a single qubit},\
  }\href@noop {} {\ }\Eprint {https://arxiv.org/abs/2210.06103} {2210.06103}
  \BibitemShut {NoStop}%
\end{thebibliography}%

\end{document}